\documentstyle[amssymb,a4,sw20lart]{article}

\makeatletter
\newcount\@hour\newcount\@minute\chardef\@x10\chardef\@xv60
\def\tcitime{
\def\@time{%
  \@minute\time\@hour\@minute\divide\@hour\@xv
  \ifnum\@hour<\@x 0\fi\the\@hour:%
  \multiply\@hour\@xv\advance\@minute-\@hour
  \ifnum\@minute<\@x 0\fi\the\@minute
  }}%

\@ifundefined{hyperref}{}{}

\@ifundefined{qExtProgCall}{\def\qExtProgCall#1#2#3#4#5#6{\relax}}{}
\def\QCTOpt[#1]#2{%
  \def\QCTOptB{#1}
  \def\QCTOptA{#2}
}
\def\QCTNOpt#1{%
  \def\QCTOptA{#1}
  \let\QCTOptB\empty
}
\def\Qct{%
  \@ifnextchar[{%
    \QCTOpt}{\QCTNOpt}
}
\def\QCBOpt[#1]#2{%
  \def\QCBOptB{#1}
  \def\QCBOptA{#2}
}
\def\QCBNOpt#1{%
  \def\QCBOptA{#1}
  \let\QCBOptB\empty
}
\def\Qcb{%
  \@ifnextchar[{%
    \QCBOpt}{\QCBNOpt}
}
\def\PrepCapArgs{%
  \ifx\QCBOptA\empty
    \ifx\QCTOptA\empty
      {}%
    \else
      \ifx\QCTOptB\empty
        {\QCTOptA}%
      \else
        [\QCTOptB]{\QCTOptA}%
      \fi
    \fi
  \else
    \ifx\QCBOptA\empty
      {}%
    \else
      \ifx\QCBOptB\empty
        {\QCBOptA}%
      \else
        [\QCBOptB]{\QCBOptA}%
      \fi
    \fi
  \fi
}
\newcount\GRAPHICSTYPE
\GRAPHICSTYPE=\z@
\def\GRAPHICSPS#1{%
 \ifcase\GRAPHICSTYPE
   \special{ps: #1}%
 \or
   \special{language "PS", include "#1"}%
 \fi
}%
\def\graffile#1#2#3#4{%
    \leavevmode
    \raise -#4 \BOXTHEFRAME{%
        \hbox to #2{\raise #3\hbox to #2{\null #1\hfil}}}%
}%
\def\draftbox#1#2#3#4{%
 \leavevmode\raise -#4 \hbox{%
  \frame{\rlap{\protect\tiny #1}\hbox to #2%
   {\vrule height#3 width\z@ depth\z@\hfil}%
  }%
 }%
}%
\newcount\draft
\draft=\z@

\newif\ifwasdraft
\wasdraftfalse

\def\GRAPHIC#1#2#3#4#5{%
 \ifnum\draft=\@ne\draftbox{#2}{#3}{#4}{#5}%
  \else\graffile{#1}{#3}{#4}{#5}%
  \fi
 }%
\def\addtoLaTeXparams#1{%
    \edef\LaTeXparams{\LaTeXparams #1}}%
\newif\ifBoxFrame \BoxFramefalse
\newif\ifOverFrame \OverFramefalse
\newif\ifUnderFrame \UnderFramefalse

\def\BOXTHEFRAME#1{%
   \hbox{%
      \ifBoxFrame
         \frame{#1}%
      \else
         {#1}%
      \fi
   }%
}

\def\doFRAMEparams#1{\BoxFramefalse\OverFramefalse\UnderFramefalse\readFRAMEparams#1\end}%
\def\readFRAMEparams#1{%
 \ifx#1\end%
  \let\next=\relax
  \else
  \ifx#1i\dispkind=\z@\fi
  \ifx#1d\dispkind=\@ne\fi
  \ifx#1f\dispkind=\tw@\fi
  \ifx#1t\addtoLaTeXparams{t}\fi
  \ifx#1b\addtoLaTeXparams{b}\fi
  \ifx#1p\addtoLaTeXparams{p}\fi
  \ifx#1h\addtoLaTeXparams{h}\fi
  \ifx#1X\BoxFrametrue\fi
  \ifx#1O\OverFrametrue\fi
  \ifx#1U\UnderFrametrue\fi
  \ifx#1w
    \ifnum\draft=1\wasdrafttrue\else\wasdraftfalse\fi
    \draft=\@ne
  \fi
  \let\next=\readFRAMEparams
  \fi
 \next
 }%
\def\IFRAME#1#2#3#4#5#6{%
      \bgroup
      \let\QCTOptA\empty
      \let\QCTOptB\empty
      \let\QCBOptA\empty
      \let\QCBOptB\empty
      #6%
      \parindent=0pt%
      \leftskip=0pt
      \rightskip=0pt
      \setbox0 = \hbox{\QCBOptA}%
      \@tempdima = #1\relax
      \ifOverFrame
          \typeout{This is not implemented yet}%
          \show\HELP
      \else
         \ifdim\wd0>\@tempdima
            \advance\@tempdima by \@tempdima
            \ifdim\wd0 >\@tempdima
               \textwidth=\@tempdima
               \setbox1 =\vbox{%
                  \noindent\hbox to \@tempdima{\hfill\GRAPHIC{#5}{#4}{#1}{#2}{#3}\hfill}\\%
                  \noindent\hbox to \@tempdima{\parbox[b]{\@tempdima}{\QCBOptA}}%
               }%
               \wd1=\@tempdima
            \else
               \textwidth=\wd0
               \setbox1 =\vbox{%
                 \noindent\hbox to \wd0{\hfill\GRAPHIC{#5}{#4}{#1}{#2}{#3}\hfill}\\%
                 \noindent\hbox{\QCBOptA}%
               }%
               \wd1=\wd0
            \fi
         \else
            \ifdim\wd0>0pt
              \hsize=\@tempdima
              \setbox1 =\vbox{%
                \unskip\GRAPHIC{#5}{#4}{#1}{#2}{0pt}%
                \break
                \unskip\hbox to \@tempdima{\hfill \QCBOptA\hfill}%
              }%
              \wd1=\@tempdima
           \else
              \hsize=\@tempdima
              \setbox1 =\vbox{%
                \unskip\GRAPHIC{#5}{#4}{#1}{#2}{0pt}%
              }%
              \wd1=\@tempdima
           \fi
         \fi
         \@tempdimb=\ht1
         \advance\@tempdimb by \dp1
         \advance\@tempdimb by -#2%
         \advance\@tempdimb by #3%
         \leavevmode
         \raise -\@tempdimb \hbox{\box1}%
      \fi
      \egroup%
}%
\def\DFRAME#1#2#3#4#5{%
 \begin{center}
     \let\QCTOptA\empty
     \let\QCTOptB\empty
     \let\QCBOptA\empty
     \let\QCBOptB\empty
     \ifOverFrame 
        #5\QCTOptA\par
     \fi
     \GRAPHIC{#4}{#3}{#1}{#2}{\z@}
     \ifUnderFrame 
        \nobreak\par #5\QCBOptA
     \fi
 \end{center}%
 }%
\def\FFRAME#1#2#3#4#5#6#7{%
 \begin{figure}[#1]%
  \let\QCTOptA\empty
  \let\QCTOptB\empty
  \let\QCBOptA\empty
  \let\QCBOptB\empty
  \ifOverFrame
    #4
    \ifx\QCTOptA\empty
    \else
      \ifx\QCTOptB\empty
        \caption{\QCTOptA}%
      \else
        \caption[\QCTOptB]{\QCTOptA}%
      \fi
    \fi
    \ifUnderFrame\else
      \label{#5}%
    \fi
  \else
    \UnderFrametrue%
  \fi
  \begin{center}\GRAPHIC{#7}{#6}{#2}{#3}{\z@}\end{center}%
  \ifUnderFrame
    #4
    \ifx\QCBOptA\empty
      \caption{}%
    \else
      \ifx\QCBOptB\empty
        \caption{\QCBOptA}%
      \else
        \caption[\QCBOptB]{\QCBOptA}%
      \fi
    \fi
    \label{#5}%
  \fi
  \end{figure}%
 }%
\newcount\dispkind%

\def\makeactives{
  \catcode`\"=\active
  \catcode`\;=\active
  \catcode`\:=\active
  \catcode`\'=\active
  \catcode`\~=\active
}
\bgroup
   \makeactives
   \gdef\activesoff{%
      \def"{\string"}
      \def;{\string;}
      \def:{\string:}
      \def'{\string'}
      \def~{\string~}
    }
\egroup

\def\FRAME#1#2#3#4#5#6#7#8{%
 \bgroup
 \@ifundefined{bbl@deactivate}{}{\activesoff}
 \ifnum\draft=\@ne
   \wasdrafttrue
 \else
   \wasdraftfalse%
 \fi
 \def\LaTeXparams{}%
 \dispkind=\z@
 \def\LaTeXparams{}%
 \doFRAMEparams{#1}%
 \ifnum\dispkind=\z@\IFRAME{#2}{#3}{#4}{#7}{#8}{#5}\else
  \ifnum\dispkind=\@ne\DFRAME{#2}{#3}{#7}{#8}{#5}\else
   \ifnum\dispkind=\tw@
    \edef\@tempa{\noexpand\FFRAME{\LaTeXparams}}%
    \@tempa{#2}{#3}{#5}{#6}{#7}{#8}%
    \fi
   \fi
  \fi
  \ifwasdraft\draft=1\else\draft=0\fi{}%
  \egroup
 }%
\def\TEXUX#1{"texux"}

\def\func#1{\mathop{\rm #1}}%
\long\def\QQQ#1#2{%
     \long\expandafter\def\csname#1\endcsname{#2}}%
\@ifundefined{QTP}{\def\QTP#1{}}{}
\@ifundefined{QEXCLUDE}{\def\QEXCLUDE#1{}}{}
\@ifundefined{Qlb}{}{}
\@ifundefined{Qlt}{}{}
\long\def\QQA#1#2{}%
\def\QTR#1#2{{\csname#1\endcsname #2}}
\def\EXPAND#1[#2]#3{}%
\def\NOEXPAND#1[#2]#3{}%
\def\LaTeXparent#1{}%
\def\ChildStyles#1{}%
\def\ChildDefaults#1{}%
\def\QTagDef#1#2#3{}%
\@ifundefined{StyleEditBeginDoc}{}{}
\def\QQfnmark#1{\footnotemark}

\@ifundefined{INDEX}{\def\INDEX#1#2{}{}}{}%
\@ifundefined{SUBINDEX}{\def\SUBINDEX#1#2#3{}{}{}}{}%
\@ifundefined{initial}%
   {\def\initial#1{\bigbreak{\raggedright\large\bf #1}\kern 2\p@\penalty3000}}%
   {}%
\@ifundefined{entry}{}{}%
\@ifundefined{primary}{}{}%
\@ifundefined{secondary}{}{}%
\@ifundefined{ZZZ}{}{\makeatletter\input gnuindex.sty\makeatother\makeindex\makeatletter}%
\@ifundefined{abstract}{%
 \def\abstract{%
  \if@twocolumn
   \section*{Abstract (Not appropriate in this style!)}%
   \else \small 
   \begin{center}{\bf Abstract\vspace{-.5em}\vspace{\z@}}\end{center}%
   \quotation 
   \fi
  }%
 }{%
 }%
\@ifundefined{endabstract}{\def\endabstract
  {\if@twocolumn\else\endquotation\fi}}{}%
\@ifundefined{maketitle}{\def\maketitle#1{}}{}%
\@ifundefined{affiliation}{\def\affiliation#1{}}{}%
\@ifundefined{proof}{}{}%
\@ifundefined{endproof}{}{}%
\@ifundefined{newfield}{\def\newfield#1#2{}}{}%
\@ifundefined{chapter}{\def\chapter#1{\par(Chapter head:)#1\par }%
 \newcount\c@chapter}{}%
\@ifundefined{part}{\def\part#1{\par(Part head:)#1\par }}{}%
\@ifundefined{section}{\def\section#1{\par(Section head:)#1\par }}{}%
\@ifundefined{subsection}{\def\subsection#1%
 {\par(Subsection head:)#1\par }}{}%
\@ifundefined{subsubsection}{\def\subsubsection#1%
 {\par(Subsubsection head:)#1\par }}{}%
\@ifundefined{paragraph}{\def\paragraph#1%
 {\par(Subsubsubsection head:)#1\par }}{}%
\@ifundefined{subparagraph}{\def\subparagraph#1%
 {\par(Subsubsubsubsection head:)#1\par }}{}%
\@ifundefined{therefore}{}{}%
\@ifundefined{backepsilon}{}{}%
\@ifundefined{yen}{}{}%
\@ifundefined{registered}{%
   \def\registered{\relax\ifmmode{}\r@gistered
                    \else$\m@th\r@gistered$\fi}%
 \def\r@gistered{^{\ooalign
  {\hfil\raise.07ex\hbox{$\scriptstyle\rm\text{R}$}\hfil\crcr
  \mathhexbox20D}}}}{}%
\@ifundefined{Eth}{}{}%
\@ifundefined{eth}{}{}%
\@ifundefined{Thorn}{}{}%
\@ifundefined{thorn}{}{}%
\@ifundefined{degree}{}{}%
\newdimen\theight
\def\Column{%
 \vadjust{\setbox\z@=\hbox{\scriptsize\quad\quad tcol}%
  \theight=\ht\z@\advance\theight by \dp\z@\advance\theight by \lineskip
  \kern -\theight \vbox to \theight{%
   \rightline{\rlap{\box\z@}}%
   \vss
   }%
  }%
 }%
\def\qed{%
 \ifhmode\unskip\nobreak\fi\ifmmode\ifinner\else\hskip5\p@\fi\fi
 \hbox{\hskip5\p@\vrule width4\p@ height6\p@ depth1.5\p@\hskip\p@}%
 }%
\def\miss{\hbox{\vrule height2\p@ width 2\p@ depth\z@}}%
\def\tcol#1{{\baselineskip=6\p@ \vcenter{#1}} \Column}  %
\def\newfmtname{LaTeX2e}
\def\chkcompat{%
   \if@compatibility
   \else
     \usepackage{latexsym}
   \fi
}

\ifx\fmtname\newfmtname
  \DeclareOldFontCommand{\rm}{\normalfont\rmfamily}{\mathrm}
  \DeclareOldFontCommand{\sf}{\normalfont\sffamily}{\mathsf}
  \DeclareOldFontCommand{\tt}{\normalfont\ttfamily}{\mathtt}
  \DeclareOldFontCommand{\bf}{\normalfont\bfseries}{\mathbf}
  \DeclareOldFontCommand{\it}{\normalfont\itshape}{\mathit}
  \DeclareOldFontCommand{\sl}{\normalfont\slshape}{\@nomath\sl}
  \DeclareOldFontCommand{\sc}{\normalfont\scshape}{\@nomath\sc}
  \chkcompat
\fi

\def\alpha{{\Greekmath 010B}}%
\def\beta{{\Greekmath 010C}}%
\def\gamma{{\Greekmath 010D}}%
\def\delta{{\Greekmath 010E}}%
\def\epsilon{{\Greekmath 010F}}%
\def\zeta{{\Greekmath 0110}}%
\def\eta{{\Greekmath 0111}}%
\def\theta{{\Greekmath 0112}}%
\def\iota{{\Greekmath 0113}}%
\def\kappa{{\Greekmath 0114}}%
\def\lambda{{\Greekmath 0115}}%
\def\mu{{\Greekmath 0116}}%
\def\nu{{\Greekmath 0117}}%
\def\xi{{\Greekmath 0118}}%
\def\pi{{\Greekmath 0119}}%
\def\rho{{\Greekmath 011A}}%
\def\sigma{{\Greekmath 011B}}%
\def\tau{{\Greekmath 011C}}%
\def\upsilon{{\Greekmath 011D}}%
\def\phi{{\Greekmath 011E}}%
\def\chi{{\Greekmath 011F}}%
\def\psi{{\Greekmath 0120}}%
\def\omega{{\Greekmath 0121}}%
\def\varepsilon{{\Greekmath 0122}}%
\def\vartheta{{\Greekmath 0123}}%
\def\varpi{{\Greekmath 0124}}%
\def\varrho{{\Greekmath 0125}}%
\def\varsigma{{\Greekmath 0126}}%
\def\varphi{{\Greekmath 0127}}%

\def\nabla{{\Greekmath 0272}}
\def\FindBoldGroup{%
   {\setbox0=\hbox{$\mathbf{x\global\edef\theboldgroup{\the\mathgroup}}$}}%
}

\def\Greekmath#1#2#3#4{%
    \if@compatibility
        \ifnum\mathgroup=\symbold
           \mathchoice{\mbox{\boldmath$\displaystyle\mathchar"#1#2#3#4$}}%
                      {\mbox{\boldmath$\textstyle\mathchar"#1#2#3#4$}}%
                      {\mbox{\boldmath$\scriptstyle\mathchar"#1#2#3#4$}}%
                      {\mbox{\boldmath$\scriptscriptstyle\mathchar"#1#2#3#4$}}%
        \else
           \mathchar"#1#2#3#4%
        \fi 
    \else 
        \FindBoldGroup
        \ifnum\mathgroup=\theboldgroup 
           \mathchoice{\mbox{\boldmath$\displaystyle\mathchar"#1#2#3#4$}}%
                      {\mbox{\boldmath$\textstyle\mathchar"#1#2#3#4$}}%
                      {\mbox{\boldmath$\scriptstyle\mathchar"#1#2#3#4$}}%
                      {\mbox{\boldmath$\scriptscriptstyle\mathchar"#1#2#3#4$}}%
        \else
           \mathchar"#1#2#3#4%
        \fi     	    
	  \fi}

\newif\ifGreekBold  \GreekBoldfalse
\let\SAVEPBF=\pbf
\def\pbf{\GreekBoldtrue\SAVEPBF}%

\@ifundefined{theorem}{}{}
\@ifundefined{lemma}{}{}
\@ifundefined{corollary}{}{}
\@ifundefined{conjecture}{}{}
\@ifundefined{proposition}{}{}
\@ifundefined{axiom}{}{}
\@ifundefined{remark}{}{}
\@ifundefined{example}{}{}
\@ifundefined{exercise}{}{}
\@ifundefined{definition}{}{}

\@ifundefined{mathletters}{%
  \newcounter{equationnumber}  
  \def\mathletters{%
     \addtocounter{equation}{1}
     \edef\@currentlabel{\theequation}%
     \setcounter{equationnumber}{\c@equation}
     \setcounter{equation}{0}%
     \edef\theequation{\@currentlabel\noexpand\alph{equation}}%
  }
  
}{}

\@ifundefined{BibTeX}{%
    \def\BibTeX{{\rm B\kern-.05em{\sc i\kern-.025em b}\kern-.08em
                 T\kern-.1667em\lower.7ex\hbox{E}\kern-.125emX}}}{}%
\@ifundefined{AmS}%
    {\def\AmS{{\protect\usefont{OMS}{cmsy}{m}{n}%
                A\kern-.1667em\lower.5ex\hbox{M}\kern-.125emS}}}{}%
\@ifundefined{AmSTeX}{}{}%
\ifx\ds@amstex\relax
\makeatother\endinput
\else
   \@ifpackageloaded{amstex}%
      {\message{amstex already loaded}\makeatother\endinput}
      {}
   \@ifpackageloaded{amsgen}%
      {\message{amsgen already loaded}\makeatother\endinput}
      {}
\fi
\let\DOTSI\relax
\def\RIfM@{\relax\ifmmode}%
\def\FN@{\futurelet\next}%
\newcount\intno@
\def\iint{\DOTSI\intno@\tw@\FN@\ints@}%
\def\iiint{\DOTSI\intno@\thr@@\FN@\ints@}%
\def\iiiint{\DOTSI\intno@4 \FN@\ints@}%
\def\idotsint{\DOTSI\intno@\z@\FN@\ints@}%
\def\ints@{\findlimits@\ints@@}%
\newif\iflimtoken@
\newif\iflimits@
\def\findlimits@{\limtoken@true\ifx\next\limits\limits@true
 \else\ifx\next\nolimits\limits@false\else
 \limtoken@false\ifx\ilimits@\nolimits\limits@false\else
 \ifinner\limits@false\else\limits@true\fi\fi\fi\fi}%
\def\multint@{\int\ifnum\intno@=\z@\intdots@                          
 \else\intkern@\fi                                                    
 \ifnum\intno@>\tw@\int\intkern@\fi                                   
 \ifnum\intno@>\thr@@\int\intkern@\fi                                 
 \int}
\def\multintlimits@{\intop\ifnum\intno@=\z@\intdots@\else\intkern@\fi
 \ifnum\intno@>\tw@\intop\intkern@\fi
 \ifnum\intno@>\thr@@\intop\intkern@\fi\intop}%
\def\intic@{%
    \mathchoice{\hskip.5em}{\hskip.4em}{\hskip.4em}{\hskip.4em}}%
\def\negintic@{\mathchoice
 {\hskip-.5em}{\hskip-.4em}{\hskip-.4em}{\hskip-.4em}}%
\def\ints@@{\iflimtoken@                                              
 \def\ints@@@{\iflimits@\negintic@
   \mathop{\intic@\multintlimits@}\limits                             
  \else\multint@\nolimits\fi                                          
  \eat@}
 \else                                                                
 \def\ints@@@{\iflimits@\negintic@
  \mathop{\intic@\multintlimits@}\limits\else
  \multint@\nolimits\fi}\fi\ints@@@}%
\def\intkern@{\mathchoice{\!\!\!}{\!\!}{\!\!}{\!\!}}%
\def\plaincdots@{\mathinner{\cdotp\cdotp\cdotp}}%
\def\intdots@{\mathchoice{\plaincdots@}%
 {{\cdotp}\mkern1.5mu{\cdotp}\mkern1.5mu{\cdotp}}%
 {{\cdotp}\mkern1mu{\cdotp}\mkern1mu{\cdotp}}%
 {{\cdotp}\mkern1mu{\cdotp}\mkern1mu{\cdotp}}}%
\def\RIfM@{\relax\protect\ifmmode}
\def\text{\RIfM@\expandafter\text@\else\expandafter\mbox\fi}
\let\nfss@text\text
\def\text@#1{\mathchoice
   {\textdef@\displaystyle\f@size{#1}}%
   {\textdef@\textstyle\tf@size{\firstchoice@false #1}}%
   {\textdef@\textstyle\sf@size{\firstchoice@false #1}}%
   {\textdef@\textstyle \ssf@size{\firstchoice@false #1}}%
   \glb@settings}

\def\textdef@#1#2#3{\hbox{{%
                    \everymath{#1}%
                    \let\f@size#2\selectfont
                    #3}}}
\newif\iffirstchoice@
\firstchoice@true
\def\Let@{\relax\iffalse{\fi\let\\=\cr\iffalse}\fi}%
\def\vspace@{\def\vspace##1{\crcr\noalign{\vskip##1\relax}}}%
\def\multilimits@{\bgroup\vspace@\Let@
 \baselineskip\fontdimen10 \scriptfont\tw@
 \advance\baselineskip\fontdimen12 \scriptfont\tw@
 \lineskip\thr@@\fontdimen8 \scriptfont\thr@@
 \lineskiplimit\lineskip
 \vbox\bgroup\ialign\bgroup\hfil$\m@th\scriptstyle{##}$\hfil\crcr}%
\def\Sb{_\multilimits@}%
\def\endSb{\crcr\egroup\egroup\egroup}%
\def\Sp{^\multilimits@}%

\newdimen\ex@
\def\rightarrowfill@#1{$#1\m@th\mathord-\mkern-6mu\cleaders
 \hbox{$#1\mkern-2mu\mathord-\mkern-2mu$}\hfill
 \mkern-6mu\mathord\rightarrow$}%
\def\leftarrowfill@#1{$#1\m@th\mathord\leftarrow\mkern-6mu\cleaders
 \hbox{$#1\mkern-2mu\mathord-\mkern-2mu$}\hfill\mkern-6mu\mathord-$}%
\def\leftrightarrowfill@#1{$#1\m@th\mathord\leftarrow
\mkern-6mu\cleaders
 \hbox{$#1\mkern-2mu\mathord-\mkern-2mu$}\hfill
 \mkern-6mu\mathord\rightarrow$}%
\def\overrightarrow{\mathpalette\overrightarrow@}%
\def\overrightarrow@#1#2{\vbox{\ialign{##\crcr\rightarrowfill@#1\crcr
 \noalign{\kern-\ex@\nointerlineskip}$\m@th\hfil#1#2\hfil$\crcr}}}%

\def\overleftarrow{\mathpalette\overleftarrow@}%
\def\overleftarrow@#1#2{\vbox{\ialign{##\crcr\leftarrowfill@#1\crcr
 \noalign{\kern-\ex@\nointerlineskip}$\m@th\hfil#1#2\hfil$\crcr}}}%
\def\overleftrightarrow{\mathpalette\overleftrightarrow@}%
\def\overleftrightarrow@#1#2{\vbox{\ialign{##\crcr
   \leftrightarrowfill@#1\crcr
 \noalign{\kern-\ex@\nointerlineskip}$\m@th\hfil#1#2\hfil$\crcr}}}%
\def\underrightarrow{\mathpalette\underrightarrow@}%
\def\underrightarrow@#1#2{\vtop{\ialign{##\crcr$\m@th\hfil#1#2\hfil
  $\crcr\noalign{\nointerlineskip}\rightarrowfill@#1\crcr}}}%

\def\underleftarrow{\mathpalette\underleftarrow@}%
\def\underleftarrow@#1#2{\vtop{\ialign{##\crcr$\m@th\hfil#1#2\hfil
  $\crcr\noalign{\nointerlineskip}\leftarrowfill@#1\crcr}}}%
\def\underleftrightarrow{\mathpalette\underleftrightarrow@}%
\def\underleftrightarrow@#1#2{\vtop{\ialign{##\crcr$\m@th
  \hfil#1#2\hfil$\crcr
 \noalign{\nointerlineskip}\leftrightarrowfill@#1\crcr}}}%

\def\qopnamewl@#1{\mathop{\operator@font#1}\nlimits@}
\let\nlimits@\displaylimits
\def\setboxz@h{\setbox\z@\hbox}

\def\varlim@#1#2{\mathop{\vtop{\ialign{##\crcr
 \hfil$#1\m@th\operator@font lim$\hfil\crcr
 \noalign{\nointerlineskip}#2#1\crcr
 \noalign{\nointerlineskip\kern-\ex@}\crcr}}}}

 \def\rightarrowfill@#1{\m@th\setboxz@h{$#1-$}\ht\z@\z@
  $#1\copy\z@\mkern-6mu\cleaders
  \hbox{$#1\mkern-2mu\box\z@\mkern-2mu$}\hfill
  \mkern-6mu\mathord\rightarrow$}
\def\leftarrowfill@#1{\m@th\setboxz@h{$#1-$}\ht\z@\z@
  $#1\mathord\leftarrow\mkern-6mu\cleaders
  \hbox{$#1\mkern-2mu\copy\z@\mkern-2mu$}\hfill
  \mkern-6mu\box\z@$}

\def\projlim{\qopnamewl@{proj\,lim}}
\def\injlim{\qopnamewl@{inj\,lim}}
\def\varinjlim{\mathpalette\varlim@\rightarrowfill@}
\def\varprojlim{\mathpalette\varlim@\leftarrowfill@}
\def\varliminf{\mathpalette\varliminf@{}}
\def\varliminf@#1{\mathop{\underline{\vrule\@depth.2\ex@\@width\z@
   \hbox{$#1\m@th\operator@font lim$}}}}
\def\varlimsup{\mathpalette\varlimsup@{}}
\def\varlimsup@#1{\mathop{\overline
  {\hbox{$#1\m@th\operator@font lim$}}}}

\def\tfrac#1#2{{\textstyle {#1 \over #2}}}%
\def\dfrac#1#2{{\displaystyle {#1 \over #2}}}%
\def\stackunder#1#2{\mathrel{\mathop{#2}\limits_{#1}}}%
\begingroup \catcode `|=0 \catcode `[= 1
\catcode`]=2 \catcode `\{=12 \catcode `\}=12
\catcode`\\=12 
|gdef|@alignverbatim#1\end{align}[#1|end[align]]
|gdef|@salignverbatim#1\end{align*}[#1|end[align*]]

|gdef|@alignatverbatim#1\end{alignat}[#1|end[alignat]]
|gdef|@salignatverbatim#1\end{alignat*}[#1|end[alignat*]]

|gdef|@xalignatverbatim#1\end{xalignat}[#1|end[xalignat]]
|gdef|@sxalignatverbatim#1\end{xalignat*}[#1|end[xalignat*]]

|gdef|@gatherverbatim#1\end{gather}[#1|end[gather]]
|gdef|@sgatherverbatim#1\end{gather*}[#1|end[gather*]]

|gdef|@gatherverbatim#1\end{gather}[#1|end[gather]]
|gdef|@sgatherverbatim#1\end{gather*}[#1|end[gather*]]

|gdef|@multilineverbatim#1\end{multiline}[#1|end[multiline]]
|gdef|@smultilineverbatim#1\end{multiline*}[#1|end[multiline*]]

|gdef|@arraxverbatim#1\end{arrax}[#1|end[arrax]]
|gdef|@sarraxverbatim#1\end{arrax*}[#1|end[arrax*]]

|gdef|@tabulaxverbatim#1\end{tabulax}[#1|end[tabulax]]
|gdef|@stabulaxverbatim#1\end{tabulax*}[#1|end[tabulax*]]

|endgroup

\def\align{\@verbatim \frenchspacing\@vobeyspaces \@alignverbatim
You are using the "align" environment in a style in which it is not defined.}

\@namedef{align*}{\@verbatim\@salignverbatim
You are using the "align*" environment in a style in which it is not defined.}
\expandafter\let\csname endalign*\endcsname =\endtrivlist

\def\alignat{\@verbatim \frenchspacing\@vobeyspaces \@alignatverbatim
You are using the "alignat" environment in a style in which it is not defined.}

\@namedef{alignat*}{\@verbatim\@salignatverbatim
You are using the "alignat*" environment in a style in which it is not defined.}
\expandafter\let\csname endalignat*\endcsname =\endtrivlist

\def\xalignat{\@verbatim \frenchspacing\@vobeyspaces \@xalignatverbatim
You are using the "xalignat" environment in a style in which it is not defined.}

\@namedef{xalignat*}{\@verbatim\@sxalignatverbatim
You are using the "xalignat*" environment in a style in which it is not defined.}
\expandafter\let\csname endxalignat*\endcsname =\endtrivlist

\def\gather{\@verbatim \frenchspacing\@vobeyspaces \@gatherverbatim
You are using the "gather" environment in a style in which it is not defined.}

\@namedef{gather*}{\@verbatim\@sgatherverbatim
You are using the "gather*" environment in a style in which it is not defined.}
\expandafter\let\csname endgather*\endcsname =\endtrivlist

\def\multiline{\@verbatim \frenchspacing\@vobeyspaces \@multilineverbatim
You are using the "multiline" environment in a style in which it is not defined.}

\@namedef{multiline*}{\@verbatim\@smultilineverbatim
You are using the "multiline*" environment in a style in which it is not defined.}
\expandafter\let\csname endmultiline*\endcsname =\endtrivlist

\def\arrax{\@verbatim \frenchspacing\@vobeyspaces \@arraxverbatim
You are using a type of "array" construct that is only allowed in AmS-LaTeX.}

\def\tabulax{\@verbatim \frenchspacing\@vobeyspaces \@tabulaxverbatim
You are using a type of "tabular" construct that is only allowed in AmS-LaTeX.}

\@namedef{arrax*}{\@verbatim\@sarraxverbatim
You are using a type of "array*" construct that is only allowed in AmS-LaTeX.}
\expandafter\let\csname endarrax*\endcsname =\endtrivlist

\@namedef{tabulax*}{\@verbatim\@stabulaxverbatim
You are using a type of "tabular*" construct that is only allowed in AmS-LaTeX.}
\expandafter\let\csname endtabulax*\endcsname =\endtrivlist

\def\@@eqncr{\let\@tempa\relax
    \ifcase\@eqcnt \def\@tempa{& & &}\or \def\@tempa{& &}%
      \else \def\@tempa{&}\fi
     \@tempa
     \if@eqnsw
        \iftag@
           \@taggnum
        \else
           \@eqnnum\stepcounter{equation}%
        \fi
     \fi
     \global\tag@false
     \global\@eqnswtrue
     \global\@eqcnt\z@\cr}

 \def\endequation{%
     \ifmmode\ifinner 
      \iftag@
        \addtocounter{equation}{-1} 
        $\hfil
           \displaywidth\linewidth\@taggnum\egroup \endtrivlist
        \global\tag@false
        \global\@ignoretrue   
      \else
        $\hfil
           \displaywidth\linewidth\@eqnnum\egroup \endtrivlist
        \global\tag@false
        \global\@ignoretrue 
      \fi
     \else   
      \iftag@
        \addtocounter{equation}{-1} 
        \eqno \hbox{\@taggnum}
        \global\tag@false%
        $$\global\@ignoretrue
      \else
        \eqno \hbox{\@eqnnum}
        $$\global\@ignoretrue
      \fi
     \fi\fi
 } 

 \newif\iftag@ \tag@false
 
 \def\tag{\@ifnextchar*{\@tagstar}{\@tag}}
 \def\@tag#1{%
     \global\tag@true
     \global\def\@taggnum{(#1)}}
 \def\@tagstar*#1{%
     \global\tag@true
     \global\def\@taggnum{#1}%
}

\makeatother

\begin{document}

\author{\ \ \ J. Pestieau, C. Smith and S. Trine. \and \quad \and \quad \\
{\it Universit\'{e} Catholique de Louvain,}\\
\\
{\it Chemin du Cyclotron 2, B-1348 Louvain-la-Neuve, Belgium}\quad \\
\quad \\
\quad \\
\quad}
\title{{\huge Positronium Decay : Gauge Invariance and Analyticity}\\
\quad \\
\quad \\
\quad \\
}
\date{May 4, 2001}
\maketitle

\begin{abstract}
\quad \newline
\quad \newline

The construction of positronium decay amplitudes is handled through the use
of dispersion relations. In this way, emphasis is put on basic QED
principles: gauge invariance and soft-photon limits (analyticity).

A firm grounding is given to the factorization approaches, and some
ambiguities in the spin and energy structures of the positronium
wavefunction are removed. Non-factorizable amplitudes are naturally
introduced. Their dynamics is described, especially regarding the
enforcement of gauge invariance and analyticity through delicate
interferences. The important question of the completeness of the present
theoretical predictions for the decay rates is then addressed. Indeed, some
of those non-factorizable contributions are unaccounted for by NRQED
analyses. However, it is shown that such new contributions are highly
suppressed, being of ${\cal O}\left( \alpha ^{3}\right) $.

Finally, a particular effective form factor formalism is constructed for
parapositronium, allowing a thorough analysis of binding energy effects and
analyticity implementation.

\quad \newline

\quad \newline

\quad \newline

\quad \newline

\quad \newline

\end{abstract}

\pagebreak

\section{Introduction}

Positronium is a bound state made of an electron with a positron. In this
paper, we will be interested in the ground states, parapositronium ($p$-$Ps$%
, singlet, $J=0$) and orthopositronium ($o$-$Ps$, triplet, $J=1$).
Positronium is unstable, it decays when its components annihilate into
photons. The lifetimes are quite different for the two species, because
parity and charge conjugation conservations imply that $p$-$Ps$ decays into
an even number of photons, while $o$-$Ps$ can only decay into an odd number.

Positronium was discovered in 1951 \cite{Discovery}. Since then, many
measurements of the lifetime of both species have been achieved. The most
precise to date are 
\begin{equation}
\Gamma ^{exp}\left( p\text{-}Ps\rightarrow 2\gamma \right) =7.9909\left(
17\right) \times 10^{9}\sec ^{-1}\;\text{\cite{Experm}\ \ \ \ \ \ \ \ \ \ \
\ \ \ \ \ \ \ \ \ \ \ \ \ \ \ \ \ \ \ \ \ \ \ \ \ }  \label{ExppPs}
\end{equation}
\begin{equation}
\Gamma ^{exp}\left( o\text{-}Ps\rightarrow 3\gamma \right) =\left\{ 
\begin{array}{l}
7.0398\left( 29\right) \,\,\,\mu sec^{-1}\qquad \text{Tokyo (SiO}_{2}\text{
Powder)\cite{Tokyo}} \\ 
7.0514\left( 14\right) \,\,\,\mu sec^{-1}\qquad \text{Ann Arbor (Gas)\cite
{AnnArborGas}} \\ 
7.0482\left( 16\right) \,\,\,\mu sec^{-1}\qquad \text{Ann Arbor (Vacuum)\cite
{AnnArborVac}}
\end{array}
\right.  \label{ExpoPs}
\end{equation}
Obviously, the situation is not clear for the $o$-$Ps$ lifetime, since
results are mutually exclusive. This is known as the orthopositronium
lifetime puzzle.

The corresponding lowest order theoretical predictions were first obtained
as 
\begin{mathletters}
\begin{eqnarray}
\Gamma _{p\text{-}Ps} &=&\frac{\alpha ^{5}m}{2}=8.033\times 10^{9}\sec
^{-1}\;\;\;\;\;\;\;\text{\cite{History}}  \label{OldTheor} \\
\Gamma _{o\text{-}Ps} &=&\alpha ^{6}m\frac{2\left( \pi ^{2}-9\right) }{9\pi }%
=7.211\,\,\,\mu \sec ^{-1}\;\text{\cite{OrePowell}}
\end{eqnarray}
with $\alpha $ the fine structure constant and $m$ the electron mass. A
great deal of work has been done on the calculation of radiative
corrections, which include perturbative QED corrections to a
non-relativistic treatment of the bound state wavefunction. The present
highly accurate theoretical predictions for the respective lifetimes are
(see for example \cite{Harris} to \cite{BS2}): 
\end{mathletters}
\begin{mathletters}
\begin{eqnarray}
\Gamma _{p\text{-}Ps} &=&\frac{\alpha ^{5}m}{2}\left( 1+\delta _{2\gamma
}+\delta _{4\gamma }\right) =7.989620\left( 13\right) \times 10^{9}\sec ^{-1}
\label{Theor} \\
\Gamma _{o\text{-}Ps} &=&\alpha ^{6}m\frac{2\left( \pi ^{2}-9\right) }{9\pi }%
\left( 1+\delta _{3\gamma }+\delta _{5\gamma }\right) =7.039965\left(
10\right) \,\,\,\mu \sec ^{-1}
\end{eqnarray}
The corrections to the dominant $2\gamma $ and $3\gamma $ modes are given by 
\end{mathletters}
\begin{eqnarray*}
\delta _{2\gamma } &=&-A_{p}\frac{\alpha }{\pi }+2\alpha ^{2}\ln \frac{1}{%
\alpha }+B_{p}\left( \frac{\alpha }{\pi }\right) ^{2}-\frac{3\alpha ^{3}}{%
2\pi }\ln ^{2}\frac{1}{\alpha }+C_{p}\frac{\alpha ^{3}}{\pi }\ln \frac{1}{%
\alpha } \\
&\approx &-0.0058828+0.0005242+0.0000277-0.0000045-0.0000048 \\
\delta _{3\gamma } &=&-A_{o}\frac{\alpha }{\pi }-\frac{\alpha ^{2}}{3}\ln 
\frac{1}{\alpha }+B_{o}\left( \frac{\alpha }{\pi }\right) ^{2}-\frac{3\alpha
^{3}}{2\pi }\ln ^{2}\frac{1}{\alpha }+C_{o}\frac{\alpha ^{3}}{\pi }\ln \frac{%
1}{\alpha } \\
&\approx &-0.0238939-0.0000873+0.0002402-0.0000045+0.0000034
\end{eqnarray*}
with coefficients 
\[
\begin{tabular}{lll}
$A_{p}=5-\pi ^{2}/4$ &  & $A_{o}=10.286606\left( 10\right) $ \\ 
$B_{p}=5.14\left( 30\right) $ &  & $B_{o}=44.52\left( 26\right) $ \\ 
$C_{p}=-7.919\left( 1\right) $ &  & $C_{o}=5.517\left( 1\right) $%
\end{tabular}
\]
The non-logarithmic ${\cal O}\left( \alpha ^{2}\right) $ corrections $%
B_{p,o} $ and logarithmic ${\cal O}\left( \alpha ^{3}\ln \alpha \right) $
corrections $C_{p,o}$ have been obtained only recently \cite{RecentCorr}.
The small contributions to the lifetimes from multi-photon decay modes are 
\cite{MultiPhot} 
\begin{eqnarray*}
\delta _{4\gamma } &=&0.274\left( 1\right) \left( \frac{\alpha }{\pi }%
\right) ^{2} \\
\delta _{5\gamma } &=&0.19\left( 1\right) \left( \frac{\alpha }{\pi }\right)
^{2}
\end{eqnarray*}
The four-photon mode increases the $p$-$Ps$ width by about $1\times
10^{5}\sec ^{-1}$, while the five-photon mode contributes for $7\times
10^{-6}\,\,\mu \sec ^{-1}$ to the $o$-$Ps$ width.

As can be observed, agreement between theory and experiment is good for
parapositronium, but the $B_{p}$, $C_{p}$ radiative corrections are not yet
accessible experimentally. For orthopositronium, these corrections render
the theoretical prediction still closer to the experimental measurement of
Ref. \cite{Tokyo} (anyway, beyond discriminating between experimental
results, $B_{o}$ and $C_{o}$ radiative corrections are not tested either).

Positronium is a test ground for bound state treatments in Quantum Field
Theory. The first try dates back to the $40$'s, with decay rates expressed
through a factorized formula \cite{History} 
\begin{equation}
\Gamma \left( p\text{-}Ps\rightarrow 2\gamma \right) =\left| \phi
_{o}\right| ^{2}\cdot \left( 4v_{rel}\sigma \left( e^{+}e^{-}\rightarrow
2\gamma \right) \right) _{v_{rel}\rightarrow 0}  \label{History}
\end{equation}
with $\phi _{o}$ the Schr\"{o}dinger positronium fundamental state
wavefunction at the origin, $\sigma \left( e^{+}e^{-}\rightarrow 2\gamma
\right) $ the total cross section for $e^{+}e^{-}\rightarrow 2\gamma $ and $%
v_{rel}$ the relative velocity of $e^{+}$ and $e^{-}$ in their
center-of-mass frame. Since then, more sophisticated decay amplitudes have
been constructed, and systematic procedures for calculating corrections have
been developed (like non-relativistic QED (NRQED), see for example \cite
{NRQED}), but factorization remains central. To avoid misunderstanding, let
us stress that by factorization is usually meant a positronium decay
amplitude built as a convolution integral, 
\begin{equation}
{\cal M}\left( p\text{-}Ps\rightarrow 2\gamma \right) \sim \int d^{3}{\bf k\;%
}\psi \left( {\bf k}\right) \cdot {\cal M}\left( e^{+}\left( -{\bf k}\right)
e^{-}\left( {\bf k}\right) \rightarrow 2\gamma \right)  \label{STD}
\end{equation}
with $\psi \left( {\bf k}\right) $ the momentum Schr\"{o}dinger wavefunction
and ${\cal M}\left( e^{+}e^{-}\rightarrow 2\gamma \right) $ the scattering
amplitude for on-shell electron-positron into photons.

Several questions concerning this formalism need to be addressed:

\begin{quote}
-- {\it The basic factorization} of the bound state dynamics from the
annihilation process remains as the basic postulate. For low order
corrections, this approximation is unquestionable, but for ${\cal O}\left(
\alpha ^{2}\right) $ corrections, factorization has to be tested. Indeed,
non-perturbative phenomena responsible for the off-shellness of the electron
and positron inside the positronium are of ${\cal O}\left( \alpha
^{2}\right) $ (since positronium mass minus twice the electron mass is of
that order). In other words, to get a sensible theoretical prediction at $%
{\cal O}\left( \alpha ^{2}\right) $, one must carefully analyze how binding
energy effects enter the general factorization approach.

-- {\it Violation of energy conservation ?} Positronium being a bound state, 
$M<2m$ ($M$ the positronium mass and $m$ the electron mass). The total
energy of the on-shell electron-positron pair entering the scattering
process is $2\sqrt{{\bf k}^{2}+m^{2}}$, clearly greater than $M$, and
getting worse as ${\bf k}$ increases in the integration.

-- The enforcement of {\it gauge invariance} may be problematic. This is
more technical. Let us say that, in general, one has to project the
electron-positron pair into a spin state compatible with the total spin of
the bound state. This bound state spin state is usually treated
non-relativistically, introducing some potential threads to the gauge
invariance of the decay amplitude.

-- Some further problems concern the {\it Bethe-Salpeter wavefunction}.
There are a lot of technicalities and subtleties associated with the bound
state wavefunction. This wavefunction can in principle be obtained
covariantly from the Bethe-Salpeter (BS) equation. Usually, one has to
introduce some approximations to solve this equation, ending with a
non-covariant four-dimensional wavefunction. This $4D$ wavefunction is then
reduced to $3D$ by making additional approximations for its temporal part
(energy-dependent part). This last reduction may be questionable. (and, as
said above, non-covariant wavefunctions can be difficult to accommodate with
gauge invariance).

-- Last but not least, {\it analytical behavior in the soft photon limit}
may be ill-defined. By this we mean that since factorization treats
intermediate charged states as on-shell, infrared singularities are
introduced. This problem is well-known, but some confusion seems to creep in
the literature. We stress that: {\bf For a model to be correct, it is not
sufficient to show that IR divergences are successfully disposed of}. From
general quantum field theory principles, a decay amplitude involving only
neutral external bosons has a correct analytical behaviour if and only if
this amplitude {\it vanishes} in the soft photon limit (not simply be
finite). This is a very profound principle, at the root of electrodynamics 
\cite{Low}.
\end{quote}

The purpose of the present paper is to answer those questions using
dispersion relations (DR). As will become clear in the course of the
presentation, DR appear as the natural framework to study positronium decay
amplitudes. Some questions on the applicability of DR analyses to bound
state decays could be raised, but, in our view, the fact that all the
questions above find consistent answers gives great confidence in this
approach.

The most prominent achievements of DR analyses are

\begin{quote}
1. a well-defined systematic procedure for factorizing the bound state
dynamics from the annihilation process,

2. an identification of the dynamics ensuring both gauge invariance and the
correct soft photon limits.
\end{quote}

The basic idea of our approach is to consider the positronium decay
amplitudes through a fully relativistic loop model, where only off-shell
constituents appear. The nice surprise is that standard three-dimensional
convolution-type amplitudes like (\ref{STD}) are recovered through a DR.
Contrary to the usual approaches, no approximations are needed to reach the
three-dimensional form. Most importantly, we will show that many variants of
the standard formula involve some unnecessary and dangerous (i.e.
gauge-dependent) approximations in their treatment of spins, and we will
remove them. Since our derivation relies on well-established techniques of
quantum field theory, we conclude that some ${\cal O}\left( \alpha
^{2}\right) $ corrections have been forgotten in those works.

It is when analytical properties are analyzed that our procedure gains its
full respectability. We will see that DR introduce some additional
non-factorizable (i.e. non-convolution-type) contributions to the decay
amplitudes. Further, some bound state structure-dependent non-factorizable
contributions must be introduced. All those additional contributions
interfere with the standard amplitude contributions to enforce both gauge
invariance and the correct analytical properties. Interestingly, because the
present approach relies on a dynamical characterization of the various
contributions, it permits an evaluation of their magnitudes. This is an
important fact because if those additional contributions prove to have been
missed in current calculations, the positronium lifetime ${\cal O}\left(
\alpha ^{2}\right) $ corrections would be wrong. We will not be able to
provide a definite answer to this question, but nevertheless point towards
an optimistic view. Provided no pathological enhancement occurs, the NRQED
result indeed should contain all ${\cal O}\left( \alpha ^{2}\right) $
corrections.

The paper is articulated in three parts. The first part contains the
presentation of the general results, and discussions about their
implications. After a short review of the standard approaches to the
construction of positronium decay amplitudes, we go on to prove that our
model reproduces those standard decay amplitudes. This result is completely
general. We then apply the insight gained to the analysis of soft-photon
behavior, and find the mechanism at play to ensure the correct analytical
limits.

The remaining of the paper contains applications and illustrations. We
analyze in some details parapositronium decay into two photons in the second
part of the paper. This decay is a bit special because the kinematics
forbids soft photon energies to vanish. Also, the appearance of a
pseudoscalar coupling (parapositronium is a pseudoscalar) has interesting
consequences, allowing form factor-like dispersion relations to be built.
This technique is then used in various calculations. This section closes by
the description of an improved basis for the perturbation series (\ref{Theor}%
), where binding energy effects are singled out.

The final part contains discussions about soft-photon limits. We take as an
example the decay of paradimuonium (the $\mu ^{+}\mu ^{-}$ bound state, $p$-$%
Dm$) into $e^{+}e^{-}\gamma $. This is the simplest positronium-like decay
process where a photon can have a vanishing energy. Again, the pseudoscalar
property of paradimuonium simplifies the discussion, allowing a thorough
analysis of the non-factorizable contributions. We also briefly discuss the
decay $K_{S}\rightarrow e^{+}e^{-}\gamma $, modeled by an intermediate
charged pion loop. The reason is the striking similarity between this
''elementary'' particle decay and the bound state analogue $p$-$%
Dm\rightarrow e^{+}e^{-}\gamma $. For example, exactly the same mechanism is
responsible for the implementation of the correct soft-photon limits (SPL).
It is very interesting to see Quantum Field Theory at play in transforming
soft-photon singularities from intermediate charged states into a {\it %
vanishing} SPL, no matter the details of the specific models. Also, we hope
that this incursion into the realm of kaon physics will convince the most
sceptical readers that our approach is viable and correct, as it is based on
general quantum field theory principles.

NB : This paper has emerged from three previous works, Ref. \cite
{PreviousWrok}.

\section{Positronium Decay Amplitudes \newline
and Dispersion Relations}

Standard amplitudes for the decay of parapositronium ($J=0$) or
orthopositronium ($J=1$) are defined as (see for example \cite{Adkins}, \cite
{RecentCorr}, \cite{TextBook1}) 
\begin{eqnarray}
{\cal M}\left( ^{2J+1}S\rightarrow A\right) &=&\sqrt{2M}\int \frac{d^{3}{\bf %
k}}{\left( 2\pi \right) ^{3}}\psi \left( {\bf k}\right) \frac{1}{\sqrt{2E_{%
{\bf k}}}}\frac{1}{\sqrt{2E_{{\bf k}}}}  \nonumber \\
&&\;\;\;\;\;\;\;\;\;\times {\cal M}\left( e^{-}\left( {\bf k},\xi \right)
,e^{+}\left( -{\bf k},\xi ^{\prime }\right) \rightarrow A\right) _{\left(
\xi ,\xi ^{\prime }\right) =\,^{2J+1}S}  \label{STD1}
\end{eqnarray}
where $M$ is the bound state mass and $\psi \left( {\bf k}\right) $ is the
Schr\"{o}dinger bound state wavefunction. This wavefunction is taken in a
convolution integral with the scattering amplitude for on-shell $%
e^{+}e^{-}\rightarrow A$, constrained to the required spin state. By working
out this constraint, we reach the commonly quoted positronium decay
amplitude (hereafter denoted 'standard approach' amplitude) 
\begin{equation}
{\cal M}\left( ^{2J+1}S\rightarrow A\right) =\sqrt{2M}\int \frac{d^{3}{\bf k}%
}{\left( 2\pi \right) ^{3}}\psi \left( {\bf k}\right) \frac{m}{2E_{{\bf k}}}%
Tr\left\{ \frac{1+\gamma ^{0}}{\sqrt{2}}{\bf P}\Gamma \left( e^{-}\left( 
{\bf k}\right) ,e^{+}\left( -{\bf k}\right) \rightarrow A\right) \right\}
\label{STD2}
\end{equation}
where 
\[
{\cal M}\left( e^{-}\left( {\bf k},\xi \right) ,e^{+}\left( -{\bf k},\xi
^{\prime }\right) \rightarrow A\right) =\overline{v}\left( -{\bf k},\xi
^{\prime }\right) \;\Gamma \left( e^{-}\left( {\bf k}\right) ,e^{+}\left( -%
{\bf k}\right) \rightarrow A\right) \;u\left( {\bf k},\xi \right) 
\]
${\bf P}=\gamma ^{5}$ for parapositronium, $%
\!\not\!\!\!\!\!\ %
e$ for orthopositronium with polarization vector $e^{\mu }$. Finally, the
width is calculated as 
\begin{equation}
\Gamma \left( ^{2J+1}S\rightarrow A\right) =\frac{1}{S}\frac{1}{2J+1}\frac{1%
}{2M}\int d\Phi _{A}\left| {\cal M}\left( ^{2J+1}S\rightarrow A\right)
\right| ^{2}  \label{DecRate}
\end{equation}
with $S$ a symmetry factor for the final state $A$.

This standard decay formula has many peculiar features, and needs a rigorous
grounding. As quoted in the introduction, factorization and SPL properties
are the most pressing problems. In section 2.1, we intend to concentrate on
the first point, leaving analytical questions to the following one. To do so
we will particularize the discussion to parapositronium decay to $2\gamma $,
which cannot suffer from any pathological SPL due to its kinematics.

Let us now introduce our model. Basically, we assume a loop structure for
the decay amplitudes. In other words, positronium decays into a virtual
electron-positron pair which subsequently annihilates into real or virtual
photons (an odd number for ortho-states, an even number for para-states).
The coupling of the positronium to its constituents is described by a form
factor, denoted by $F_{B}$. It is not assumed to be a constant, since a
constant form factor would amount to consider positronium as a point-like
bound state.

For parapositronium decay into two photons, our model is represented in
figure 1. The corresponding amplitude is 
\begin{equation}
{\cal M}\left( p\text{-}Ps\rightarrow \gamma \gamma \right) =\int \frac{%
d^{4}q}{\left( 2\pi \right) ^{4}}F_{B}Tr\left\{ \gamma _{5}\frac{i}{%
\!\not\!%
q-\frac{1}{2}%
\!\not\!%
P-m}\Gamma ^{\mu \nu }\frac{i}{%
\!\not\!%
q+\frac{1}{2}%
\!\not\!%
P-m}\right\} \varepsilon _{1\mu }^{*}\varepsilon _{2\nu }^{*}  \label{Loop1}
\end{equation}
with $P$ the positronium four-momentum and $F_{B}\equiv F_{B}\left(
q^{2},P\cdot q\right) $ the form factor. The tensor $\Gamma ^{\mu \nu }$ is
the scattering amplitude for off-shell $e^{+}e^{-}$, with incoming momenta $%
\frac{1}{2}P-q$ and $\frac{1}{2}P+q$, into two photons : 
\begin{eqnarray}
\Gamma ^{\mu \nu } &=&\Gamma ^{\mu \nu }\left( e^{+}\left( \tfrac{1}{2}%
P-q\right) e^{-}\left( \tfrac{1}{2}P+q\right) \rightarrow \gamma \left(
l_{1}\right) \gamma \left( l_{2}\right) \right)  \nonumber \\
&=&ie\gamma ^{\mu }\frac{i}{%
\!\not\!%
q-\frac{1}{2}%
\!\not\!%
P+%
\!\not\!%
l_{1}-m}ie\gamma ^{\nu }+ie\gamma ^{\nu }\frac{i}{%
\!\not\!%
q+\frac{1}{2}%
\!\not\!%
P-%
\!\not\!%
l_{1}-m}ie\gamma ^{\mu }  \label{Loop2}
\end{eqnarray}

\paragraph{Remarks :}

\begin{quote}
1) The model is extended to orthopositronium decays through the replacement
of $\gamma _{5}$ by $%
\!\not\!\!\!\ %
e$.

2 ) The electron and positron in the loop are never on-shell, because $M<2m$.

3 ) $F_{B}$ contains all the information about the bound state. Let us
postulate a form for this coupling as (in the positronium center-of-mass
frame) 
\begin{equation}
F_{B}\equiv C\phi _{0}{\cal F}\left( q_{o},{\bf q}^{2}\right) \left( {\bf q}%
^{2}+\gamma ^{2}\right)  \label{Loop3}
\end{equation}
with $C$ a constant, and $\gamma ^{2}=m^{2}-M^{2}/4$ related to the binding
energy $E_{B}=2m-M$. In QED, $E_{B}$ and $\gamma ^{2}$ are related to the
fine structure constant as $\gamma ^{2}\approx m^{2}\alpha ^{2}/4$ and $%
E_{B}=-m\alpha ^{2}/4$.
\end{quote}

\subsection{The loop model reproduces standard decay amplitudes}

We can state our result in three steps (details are found in the appendix).

First, we define 
\begin{equation}
\func{Im}{\cal T}\left( P^{2}\right) \equiv \func{Im}{\cal M}\left( p\text{-}%
Ps\left( P^{2}\right) \rightarrow \gamma \gamma \right)  \label{TH1}
\end{equation}
with the absorptive part given by the two vertical cuts, figure 2.

Then, using an unsubstracted dispersion relation (see \cite{TextBook2}, \cite
{Kniehl}) with $s=P^{2}$: 
\begin{equation}
{\cal T}\left( M^{2}\right) =\func{Re}{\cal T}\left( M^{2}\right) =\frac{1}{%
\pi }\int_{4m^{2}}^{\infty }\frac{ds}{s-M^{2}}\func{Im}{\cal T}\left(
s\right)  \label{TH2}
\end{equation}
(${\cal T}\left( M^{2}\right) =\func{Re}{\cal T}\left( M^{2}\right) $
because $M<2m$), and changing of variables, one recovers a factorized form
(i.e. a convolution type amplitude): 
\begin{equation}
{\cal M}\left( p\text{-}Ps\rightarrow \gamma \gamma \right) =\frac{C}{2}\int 
\frac{d^{3}{\bf k}}{\left( 2\pi \right) ^{3}}\frac{\phi _{0}{\cal F}\left( 0,%
{\bf k}^{2}\right) }{2E_{{\bf k}}}Tr\left\{ \gamma _{5}\left( m-%
\!\not\!\!\!\ %
k^{\prime }\right) \Gamma \left( m+%
\!\not\!\!\!\ %
k\right) \right\}  \label{TH3}
\end{equation}
with $\Gamma $ given by $\Gamma ^{\mu \nu }\left( e^{-}\left( k\right)
,e^{+}\left( k^{\prime }\right) \rightarrow \gamma \gamma \right)
\varepsilon _{1\mu }^{*}\varepsilon _{2\nu }^{*}$ and $k=\left( E_{k},{\bf k}%
\right) ,$ $k^{\prime }=\left( E_{k},-{\bf k}\right) $, i.e. the same $%
\Gamma $ as in (\ref{STD2}).

Finally, if one further identifies $C=\sqrt{M}/m$ and $\phi _{0}{\cal F}%
\left( 0,{\bf k}^{2}\right) =\psi \left( {\bf k}\right) $, by neglecting the 
${\bf k}$ dependence in the projectors $\left( m-%
\!\not\!\!\!\ %
k^{\prime }\right) \ $and $\left( m+%
\!\not\!\!\!\ %
k\right) $, (\ref{TH3}) reduces to (\ref{STD2}).

As a corollary, the energy-dependence of the form factor is demonstrated to
be irrelevant. Indeed, the dispersion relation fixes $q_{o}=0$. From now on,
we will simplify the notation and identify ${\cal F}\left( 0,{\bf k}%
^{2}\right) ={\cal F}\left( {\bf k}^{2}\right) $.

\subsubsection{Discussion}

Let us now comment on this result. Interpreting standard approach amplitudes
as dispersion relation integrals, we can give answers to most of the
questions raised in the introduction.

First, factorization appears simply as a manifestation of the optical
theorem: the appearance of on-shell intermediate states is expected in the
imaginary part. The dispersive integral ''shifts'' them off-shell: the
amplitude at the physical point $s=M^{2}$ is purely real, i.e. only
off-shell $e^{+}e^{-}$ circulate inside the loop. Also, the apparent
non-conservation of energy is explained, since the dispersive integral is
done along the loop model imaginary part cut, where the initial energy is
indeed sufficient to get on-shell constituents. In other words, the standard
approach amplitudes are constructed like in ''old-fashioned perturbation
theory''. In the context of quantum field theory, their natural framework is
dispersion theory.

Second, the spin projections are treated covariantly. The projector often
seen in the literature is an approximation. In those approaches, the
projector $\left( 1+\gamma _{0}\right) {\bf P}$ comes from the spin
wavefunction of the positronium, while here it comes directly from the loop
propagators. This is more adequate since it is the moving electron-positron
entering the scattering process that must be constrained to the required
spin state. The projectors appearing in (\ref{TH3}) for particles in motion
will introduce some new corrections to the positronium decay rate of the
order of the binding energy $\gamma ^{2}$, i.e. $\alpha ^{2}$ (here ''new''
means relatively to the result found starting from (\ref{STD2}) or any
equivalent formula).

Further, the covariant projectors have the important property of preserving
gauge invariance. Indeed, when using $\left( 1+\gamma _{0}\right) {\bf P}$
as projector, gauge invariance is valid only if the constituents are taken
at rest, i.e. ${\bf k}\equiv 0$ (the so-called static limit). In other
words, (\ref{TH3}) is a gauge invariant positronium decay amplitude while (%
\ref{STD2}) is not.

Another consequence is that since the loop model appears at the root of
standard approaches, interesting alternative calculational tools can be
built from it. In particular, it provides for an easy procedure to integrate
the wavefunction exactly, keeping tracks of binding energy effects even at
lowest order. This is the subject of section 3.

\subsubsection{A comment about Bethe-Salpeter}

Viewing the decay process through a loop model is far from new. After all,
it is the approach followed using Bethe-Salpeter (BS) analyses, where one
starts with 
\begin{equation}
{\cal M}\left( Ps\rightarrow n\gamma \right) =\int \frac{d^{4}q}{\left( 2\pi
\right) ^{4}}Tr\left\{ \Gamma \left( e^{-}e^{+}\rightarrow n\gamma \right)
\Psi \left( q\right) \right\}  \label{BSdecay}
\end{equation}
with $\Gamma $ the off-shell scattering amplitude (given in (\ref{Loop2})
for $n=2$). The loop structure emerges from the BS identification of the
wavefunction as 
\[
\Psi \left( q\right) =\frac{i}{\frac{1}{2}%
\!\not\!%
P+%
\!\not\!%
q-m}\Gamma _{BS}\left( \frac{P}{2}+q,\frac{P}{2}-q,P\right) \frac{i}{\frac{1%
}{2}%
\!\not\!%
P-%
\!\not\!%
q-m} 
\]
with $\Gamma _{BS}$ the BS vertex function. All this is standard textbook
material (see for example \cite{TextBook3}).

As is well-known, the BS equation cannot be solved exactly, and one has to
rely on some approximations. The most popular one for positronium is that of
Barbieri-Remiddi \cite{BarbRem}, which can be viewed as a four-dimensional
generalization of the usual Schr\"{o}dinger wavefunction. To make contact
with the three-dimensional integral representation for the decay amplitude (%
\ref{STD1}), the wavefunction is approximated as $\Psi \left( q\right) \sim
\delta \left( q_{o}\right) \psi \left( {\bf q}\right) $ in (\ref{BSdecay})
(see for example \cite{Adkins}, \cite{BS2}).

The novelty of our approach is to use dispersion relations to study the BS
loop. This has great advantages over approximate methods, because our
treatment is {\it exact}. Also, all the previous comments show the insights
gained doing the reduction our way. In fact, starting with a
four-dimensional BS form factor or vertex function $\Gamma _{BS}\sim
F_{B}\left( q_{0},{\bf q}\right) $, {\it the dispersion relations alone
enforce} $q_{0}=0$, i.e. the energy-dependence is set to zero.

Another particularity of our approach is the treatment of spin, since we
took the BS vertex $\Gamma _{BS}$ with its spin structure replaced by $%
\gamma _{5}$ or $%
\not\!\!\!\!\!\ %
e$ ($F_{B}\left( q_{0},{\bf q}\right) $ is a scalar), a replacement dictated
by the bound state behavior under parity and charge conjugation. This allows
us to identify the correct covariant spin projectors for the $e^{+}e^{-}$
pair, and to preserve manifest gauge invariance.

In conclusion, the present dispersion technique could help clarify some
problems concerning the reduction of four-dimensional BS wavefunctions to
three-dimensional ones (note that one should not confuse the present
reduction of the BS {\it wavefunction} with the $4D$ $\rightarrow $ $3D$
reduction of the BS {\it equation} itself, extensively discussed in the
literature, see for example \cite{Wallace} and references quoted there).

\subsection{Analytical Properties of Positronium Decay Amplitudes}

We have just seen that the standard positronium decay amplitudes are given a
natural interpretation in terms of dispersion relation integrals. Let us see
the implications for orthopositronium decay into three photons.

The loop model amplitude consists of six diagrams (figure 3) 
\begin{eqnarray}
{\cal M}\left( o\text{-}Ps\rightarrow \gamma \gamma \gamma \right)
&=&e_{\alpha }\left( P\right) \int \frac{d^{4}q}{\left( 2\pi \right) ^{4}}%
F_{B}\times \varepsilon _{1\mu }^{*}\varepsilon _{2\nu }^{*}\varepsilon
_{3\rho }^{*}  \nonumber \\
&&\times Tr\left\{ \gamma ^{\alpha }\frac{i}{%
\!\not\!%
q-\frac{1}{2}%
\!\not\!%
P-m}\Gamma ^{\mu \nu \rho }\frac{i}{%
\!\not\!%
q+\frac{1}{2}%
\!\not\!%
P-m}\right\}  \label{OrthoLoop}
\end{eqnarray}
where the off-shell scattering amplitudes are 
\begin{eqnarray}
\Gamma ^{\mu \nu \rho } &\equiv &\Gamma ^{\mu \nu \rho }\left( e^{+}(\tfrac{1%
}{2}P-q)e^{-}(\tfrac{1}{2}P+q)\rightarrow \gamma \left( l_{1}\right) \gamma
\left( l_{2}\right) \gamma \left( l_{3}\right) \right)  \label{OrthoScatt} \\
&=&-ie^{3}\left( \Gamma _{12}^{\mu \nu \rho }+\Gamma _{32}^{\rho \nu \mu
}+\Gamma _{31}^{\rho \mu \nu }\right)  \nonumber
\end{eqnarray}
with 
\begin{eqnarray*}
\Gamma _{ij}^{\mu \nu \rho } &=&\gamma ^{\nu }\frac{1}{%
\!\not\!%
q-\frac{1}{2}%
\!\not\!%
P+%
\!\not\!%
l_{j}-m}\gamma ^{\rho }\frac{1}{%
\!\not\!%
q+\frac{1}{2}%
\!\not\!%
P-%
\!\not\!%
l_{i}-m}\gamma ^{\mu } \\
&&+\gamma ^{\mu }\frac{1}{%
\!\not\!%
q-\frac{1}{2}%
\!\not\!%
P+%
\!\not\!%
l_{i}-m}\gamma ^{\rho }\frac{1}{%
\!\not\!%
q+\frac{1}{2}%
\!\not\!%
P-%
\!\not\!%
l_{j}-m}\gamma ^{\nu }
\end{eqnarray*}
and $P=l_{1}+l_{2}+l_{3}$. To compute these loop amplitudes, we start by
extracting the imaginary part, and we discover a completely different
picture than the one described earlier for the two-photon parapositronium
decay.

Since the discussion is rather involved, the following summary of what we
intend to show may be useful:

\begin{quote}
{\it 1. There are six vertical cut diagrams }${\cal V}$ (figure 4a){\it ,
with the properties:}

-- reproduce the standard convolution-type decay amplitude;

-- are separately gauge invariant;

-- are badly behaved in the soft-photon limit, i.e. violation of analyticity.

{\it 2. There are additional contributions arising from the oblique cuts }$%
{\cal D}$ (figure 4b){\it , with the properties:}

-- are not of the factorized convolution-type;

-- are not gauge invariant by themselves, except for a constant $F_{B}$;

-- also badly behaved in the IR; the combination ${\cal V}+{\cal D}$ also
fails to vanish in the SPL, except for a constant $F_{B}$.

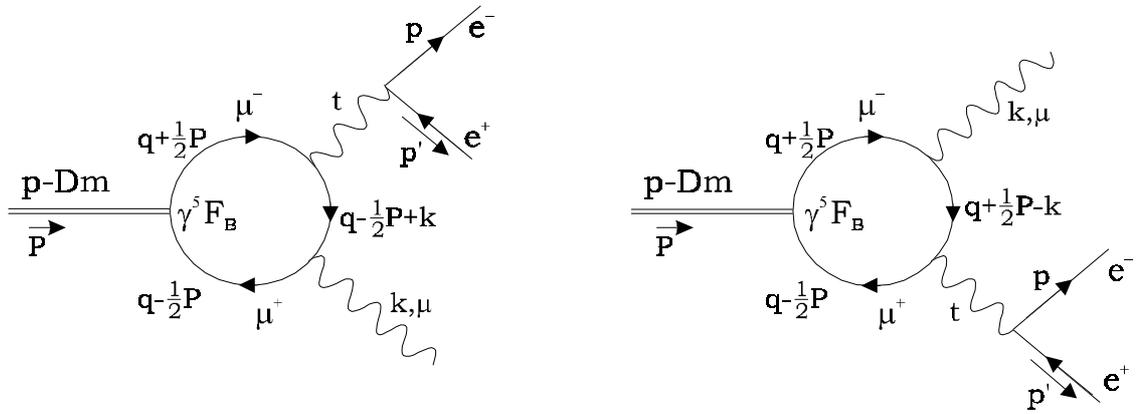
3. {\it In addition to the loop model amplitudes }${\cal V}$ and ${\cal D}$%
{\it , structure-dependent contributions }${\cal S}${\it \ must be
considered }(figure 4c){\it , with the properties:}

-- are non factorizable, requiring a new form factor for the vertex $Ps$ $%
e^{+}$ $e^{-}$ $\gamma $;

-- act to restore gauge invariance;

-- interfere with both ${\cal V}$ and ${\cal D}$ to ensure the correct
soft-photon limits.

(if $F_{B}$ is constant, structure terms need not be considered)

4. {\it The various contributions should scale as} 
\[
\frac{{\cal D}}{{\cal V}}\sim {\em O}\left( \alpha ^{2}\right) ,\;\;\frac{%
{\cal S}}{{\cal V}}\sim {\em O}\left( \alpha ^{3}\right) 
\]
\end{quote}

We will spend some time to justify and explain these results, reviewing each
point in turn. The strategy is to use dispersion relations as a
classification scheme for contributions to the decay process, and
fundamental principles (gauge invariance and soft-photon limits) to
characterize and constraint each type of contributions.

Let us state our results more precisely, introducing some notations. We
write the decay amplitude 
\begin{eqnarray*}
{\cal M}\left( o\text{-}Ps\left( M^{2}\right) \rightarrow \gamma \gamma
\gamma \right) &=&\left( {\cal M}^{\mu \nu \rho }\left( \text{loop model}%
\right) +{\cal M}^{\mu \nu \rho }\left( \text{structure}\right) \right)
\varepsilon _{1\mu }^{*}\varepsilon _{2\nu }^{*}\varepsilon _{3\rho }^{*} \\
&\equiv &\left( {\cal V}^{\mu \nu \rho }+{\cal D}^{\mu \nu \rho }+{\cal S}%
^{\mu \nu \rho }\right) \varepsilon _{1\mu }^{*}\varepsilon _{2\nu
}^{*}\varepsilon _{3\rho }^{*} \\
&\equiv &{\cal V}+{\cal D}+{\cal S}
\end{eqnarray*}
Using the optical theorem, each contribution is expressed as 
\begin{mathletters}
\begin{eqnarray}
{\cal V} &=&\frac{1}{\pi }\int_{4m^{2}}^{\infty }\frac{ds}{s-M^{2}}\func{Im}%
{\cal V}\left( s\right)  \label{DRvert} \\
{\cal D} &=&\frac{1}{\pi }\sum_{i=1}^{3}\int_{s_{\min }\left( l_{i}\right)
}^{\infty }\frac{ds}{s-M^{2}}\func{Im}{\cal D}_{i}\left( s\right)
\label{DRob} \\
{\cal S} &=&\frac{1}{\pi }\sum_{i=1}^{3}\int_{s_{\min }\left( l_{i}\right)
}^{\infty }\frac{ds}{s-M^{2}}\func{Im}{\cal S}_{i}\left( s\right)
\label{DRsd}
\end{eqnarray}
with the vertical cuts imaginary part 
\end{mathletters}
\begin{equation}
\func{Im}{\cal V}\left( P^{2}\right) =\varepsilon _{1\mu }^{*}\varepsilon
_{2\nu }^{*}\varepsilon _{3\rho }^{*}\int d\Phi _{2}{\cal M}\left( o\text{-}%
Ps\left( P^{2}\right) \rightarrow e^{+}e^{-}\right) \times {\cal M}%
_{scatt}^{\mu \nu \rho }\left( e^{+}e^{-}\rightarrow \gamma _{l_{1}}\gamma
_{l_{2}}\gamma _{l_{3}}\right)  \label{ImVert}
\end{equation}
The integration is over the two-body $e^{+}e^{-}$ phase-space $d\Phi _{2}$,
and summation over spins is understood. Oblique and structure-dependent
imaginary parts are 
\begin{equation}
\func{Im}{\cal D}_{1}\left( P^{2}\right) =\varepsilon _{1\mu
}^{*}\varepsilon _{2\nu }^{*}\varepsilon _{3\rho }^{*}\int d\Phi _{2}{\cal M}%
_{IB;1}^{\mu }\left( o\text{-}Ps\left( P^{2}\right) \rightarrow
e^{+}e^{-}\gamma _{l_{1}}\right) \times {\cal M}_{scatt;23}^{\nu \rho
}\left( e^{+}e^{-}\rightarrow \gamma _{l_{2}}\gamma _{l_{3}}\right)
\label{ImOB}
\end{equation}
\begin{equation}
\func{Im}{\cal S}_{1}\left( P^{2}\right) =\varepsilon _{1\mu
}^{*}\varepsilon _{2\nu }^{*}\varepsilon _{3\rho }^{*}\int d\Phi _{2}{\cal M}%
_{SD;1}^{\mu }\left( o\text{-}Ps\left( P^{2}\right) \rightarrow
e^{+}e^{-}\gamma _{l_{1}}\right) \times {\cal M}_{scatt;23}^{\nu \rho
}\left( e^{+}e^{-}\rightarrow \gamma _{l_{2}}\gamma _{l_{3}}\right)
\label{ImSD}
\end{equation}
and similarly for photon $2,3$. The oblique cut amplitudes ${\cal M}_{IB}$
are interpreted as bremsstrahlung amplitudes, hence the subscript.

Gauge invariance is expressed through Ward identities. Let us concentrate on
the photon $l_{1}$ for definiteness. First note that scattering amplitudes
are all gauge invariant 
\begin{eqnarray*}
l_{1,\mu }{\cal M}_{scatt}^{\mu \nu \rho }\left( e^{+}e^{-}\rightarrow
\gamma _{l_{1}}\gamma _{l_{2}}\gamma _{l_{3}}\right) &=&0 \\
l_{1,\mu }{\cal M}_{scatt;12\left( 3\right) }^{\mu \nu }\left(
e^{+}e^{-}\rightarrow \gamma _{l_{1}}\gamma _{l_{2\left( 3\right) }}\right)
&=&0
\end{eqnarray*}
because the $e^{+}e^{-}$ are on-shell. This implies gauge invariance for the
vertical cuts 
\[
l_{1,\mu }{\cal V}^{\mu \nu \rho }=0 
\]
On the other hand, $IB$ and $SD$ amplitudes cannot be separately gauge
invariant, since $l_{1,\mu }{\cal M}_{IB;1}^{\mu }\neq 0$. This serves as a
constraint for $SD$ amplitudes: we impose 
\[
l_{1,\mu }{\cal M}_{IB;1}^{\mu }+l_{1,\mu }{\cal M}_{SD;1}^{\mu }=0 
\]
Provided this last equality holds, the positronium decay amplitude is gauge
invariant.

To analyze the soft-photon behaviors, let us expand each amplitude around $%
l_{1}^{0}=0$: 
\begin{eqnarray*}
&&{\cal V}\stackrel{l_{1}^{0}\sim 0}{=}\frac{v_{1}}{l_{1}^{0}}+v_{2}+{\cal O}%
\left( l_{1}^{0}\right) \\
&&{\cal D}\stackrel{l_{1}^{0}\sim 0}{=}\frac{d_{1}}{l_{1}^{0}}+d_{2}+{\cal O}%
\left( l_{1}^{0}\right) \\
&&{\cal S}\stackrel{l_{1}^{0}\sim 0}{=}s_{2}+{\cal O}\left( l_{1}^{0}\right)
\end{eqnarray*}
Selection rules alone suffice to cancel IR divergences, i.e. $v_{1}=d_{1}=0$%
, and 
\[
{\cal M}\left( o\text{-}Ps\left( M^{2}\right) \rightarrow \gamma
_{l_{1}}\gamma _{l_{2}}\gamma _{l_{3}}\right) \stackrel{l_{1}^{0}\sim 0}{%
\sim }\text{Constant} 
\]
The implementation of the correct SPL requires in addition 
\[
v_{2}+d_{2}+s_{2}=0 
\]
which guarantees 
\[
{\cal M}\left( o\text{-}Ps\left( M^{2}\right) \rightarrow \gamma
_{l_{1}}\gamma _{l_{2}}\gamma _{l_{3}}\right) \stackrel{l_{1}^{0}\sim 0}{%
\sim }{\cal O}\left( l_{1}^{0}\right) 
\]
Similar behaviors are expected for photon $2,3$.

Let us now prove those assertions, and discuss their interpretations.

\subsubsection{Factorization properties}

By an immediate application of the result of section 2.1, one can easily
show that {\it vertical cuts alone reproduce the standard three-dimensional
convolution-type amplitude}: 
\begin{eqnarray}
{\cal V} &=&\frac{C}{2}\int \frac{d^{3}{\bf k}}{\left( 2\pi \right) ^{3}}%
\frac{\varphi _{0}{\cal F}\left( {\bf k}^{2}\right) }{2E_{{\bf k}}}%
\varepsilon _{1\mu }^{*}\varepsilon _{2\nu }^{*}\varepsilon _{3\rho }^{*} 
\nonumber \\
&&\;\;\;\;\times Tr\left\{ 
\!\not\!\!\!\ %
e\left( m-%
\!\not\!\!\!\ %
k^{\prime }\right) \Gamma ^{\mu \nu \rho }\left( e^{+}\left( k^{\prime
}\right) e^{-}\left( k\right) \rightarrow \gamma \gamma \gamma \right)
\left( m+%
\!\not\!\!\!\ %
k\right) \right\}  \label{STDOrtho}
\end{eqnarray}
with on-shell $e^{+}e^{-}$. Note that it is important to treat spin
projectors covariantly to preserve gauge invariance (if $\left( m+%
\!\not\!\!\!\ %
k\right) 
\!\not\!\!\!\ %
e\left( m-%
\!\not\!\!\!\ %
k^{\prime }\right) $ is approximated as $\left( 1+\gamma _{0}\right) 
\!\not\!\!\!\ %
e$, gauge invariance is lost, except in the static limit).

Bremsstrahlung amplitudes generate a factorization problem for oblique cuts $%
{\cal D}$. In fact, the emission of a photon by either the electron or the
positron breaks the ''symmetry'' of the diagram (by symmetry is meant a
configuration in which each lepton is carrying half the positronium
momentum). Let us recall the implications of this symmetry for the vertical
cuts. The first term in (\ref{ImVert}) is 
\begin{equation}
{\cal M}\left( o\text{-}Ps\left( P^{2}\right) \rightarrow e^{+}\left( \tfrac{%
P}{2}-q\right) e^{-}\left( \tfrac{P}{2}+q\right) \right) \sim F_{B}\left( q-%
\tfrac{1}{2}P,q+\tfrac{1}{2}P\right)  \label{SymPoint}
\end{equation}
The energy conservation at that vertex and the on-shell condition for the $%
e^{+}e^{-}$ (in the $o$-$Ps$ center-of-mass frame) amount to 
\[
\delta \left( q_{0}\right) \delta \left( \left| {\bf q}\right| -\sqrt{%
P^{2}/4-m^{2}}\right) 
\]
This allows us to write $F_{B}$ very simply as $F_{B}=C\phi _{0}{\cal F}%
\left( {\bf q}^{2}\right) \left( {\bf q}^{2}+\gamma ^{2}\right) $ (\ref
{Loop3}). Such simplifications do not occur for the oblique cuts, the
energy-dependence of the form factor is not set to zero but to a complicated 
${\bf q}$-dependent value. We refer to the calculation of oblique cuts for $%
p $-$Dm\rightarrow e^{+}e^{-}\gamma $ (section 4) for an explicit
illustration of the problem.

\subsubsection{Gauge Invariance and Low's theorem}

Low's theorem states that {\it an amplitude involving only neutral external
bosons must vanish in the soft photon limit}, i.e. when anyone photon's
energy is taken to zero.

This theorem must apply to positronium decay amplitudes. One must then
conclude that standard amplitudes do not exhibit the correct analytical
behavior, since individual scattering amplitudes contained in $\Gamma \left(
e^{-}\left( {\bf k}\right) ,e^{+}\left( -{\bf k}\right) \rightarrow \gamma
\gamma \gamma \right) $ explode in the SPL. One can view this simply by
noting the presence in $\Gamma $ of electron propagators of the form 
\[
\gamma _{\mu }\frac{i}{%
\!\not\!\!\!\ %
k-%
\!\not\!\!\!\ %
l_{i}-m}\gamma _{\nu }\;\cdots 
\]
and when the $i$th photon energy $l_{i}^{0}$ is vanishing, the propagator
explodes because the constituents are on-shell. In fact, the situation is
improved by a consequence of selection rules: when the six vertical cut
amplitudes are considered, these IR divergences cancel and ${\cal M}\left( o%
\text{-}Ps\rightarrow \gamma \gamma \gamma \right) $ is finite. To
understand this, remember that an IR divergence in ${\cal M}\left(
e^{+}e^{-}\rightarrow \gamma \gamma \gamma \right) $ would be cancelled by
IR divergences in radiative corrections to ${\cal M}\left(
e^{+}e^{-}\rightarrow \gamma \gamma \right) $, which cannot contribute to $%
{\cal M}\left( o\text{-}Ps\rightarrow \gamma \gamma \gamma \right) $.
Nevertheless, and contrary to what is often believed, selection rules do not
suffice to enforce analyticity for the standard decay amplitude, because $%
{\cal M}\left( o\text{-}Ps\rightarrow \gamma \gamma \gamma \right) $ must
vanish and not simply be finite.

Looking back at the loop model, equivalent to the Bethe-Salpeter loop, a
possible solution immediately emerges. Since only off-shell fermions can
circulate in the loop, no analytical problem should occur. Taking into
account all the cuts, both the oblique and the vertical ones, the correct
SPL should be restored. In fact, this is {\it wrong}, because the loop model
amplitude is not gauge invariant for $o$-$Ps$ decays, and Low's theorem
heavily relies on gauge invariance, i.e. Ward identities. Let us discuss
this issue of gauge invariance more precisely.

The source of the problem is in the momentum-dependence of the form factor.
One can show that the loop model amplitude is gauge invariant if and only if
the form factor accommodates linear shifts $F_{B}\left( q\right)
=F_{B}\left( q+l_{1}\right) $, and similarly for $l_{2},l_{3}$. For a
constant form factor, the loop is gauge invariant, and no problem of SPL
occurs (the loop amplitude is then obtained from the light-by-light
scattering amplitude, for which it is well-known that gauge invariance
guarantees correct SPL \cite{TextBook1}).

Equivalently, we can trace the problem down to the bremsstrahlung
amplitudes. To visualize the situation, let us change the momentum
parametrization and draw amplitudes for $o$-$Ps\rightarrow e^{+}\left(
p_{1}\right) e^{-}\left( p_{2}\right) \gamma \left( l_{1}\right) $, as shown
in figure 5, 
\begin{eqnarray}
{\cal M}_{IB;1}^{\mu } &=&ieF_{B}\left( p_{1}+l_{1},p_{2}\right) \left\{ 
\overline{u}\left( p_{1}\right) \frac{2p_{1}^{\mu }+\gamma ^{\mu }%
\!\not\!%
l_{1}}{2p_{1}\cdot l_{1}}\gamma ^{\alpha }v\left( p_{2}\right) \right\}
e_{\alpha }\left( P\right)  \nonumber \\
&&+ieF_{B}\left( p_{1},p_{2}+l_{1}\right) \left\{ \overline{u}\left(
p_{1}\right) \gamma ^{\alpha }\frac{-2p_{2}^{\mu }-%
\!\not\!%
l_{1}\gamma ^{\mu }}{2p_{2}\cdot l_{1}}v\left( p_{2}\right) \right\}
e_{\alpha }\left( P\right)  \label{Brem}
\end{eqnarray}
Obviously, when contracted by $l_{1}^{\mu }$, the two amplitudes fail to
cancel each other, due to the different momentum dependences of $F_{B}$: 
\[
l_{1,\mu }{\cal M}_{IB;1}^{\mu }=ie\left[ F_{B}\left(
p_{1}+l_{1},p_{2}\right) -F_{B}\left( p_{1},p_{2}+l_{1}\right) \right]
\left\{ \overline{u}\left( p_{1}\right) \gamma ^{\alpha }v\left(
p_{2}\right) \right\} e_{\alpha }\left( P\right) \neq 0 
\]
Thus gauge invariance is violated by the bremsstrahlung parts of the oblique
cuts.

{\it Dispersion theory permits the precise identification of the
bremsstrahlung processes as responsible for both analyticity and gauge
invariance violations. }The idea is now to add new structure-dependent (SD)
contributions, such that gauge invariance is restored. What we will show is
that these new SD contributions also automatically restore the analytical
properties for the whole amplitude.

It could seem that the loop model has not achieved much. Worse, by
considering it, we have not solved the analytical problems, and we have
brought in new gauge invariance problems (since vertical cuts are separately
gauge invariant). One could therefore think that it would be much easier to
try to find new structure-dependent gauge invariant contributions to cure
the standard amplitudes (\ref{STDOrtho}), and forget about loops.

Our point of view is that one must consider the loop model as the starting
point, being strictly equivalent to the Bethe-Salpeter approaches (\ref
{BSdecay}). Another argument is that oblique cuts do a great deal of the
work: for a constant form factor, they succeed at curing vertical cut
analytical problems. The new SD contributions are therefore accounting for
the non-constancy of the form factor. This is far more satisfactory. The
loop represents the decay dynamics, and ''nearly'' succeeds at restoring
SPL. The remaining violation comes from the bound state dynamics
(non-constant form factor), and is cured by new bound state
structure-dependent processes.

Another advantage of the loop approach is that pseudoscalar para-state
decays do not necessitate structure-dependent terms. As we will see, thanks
to the appearance of a pseudoscalar coupling $\gamma _{5}$, the loop
amplitude is gauge invariant, and well-behaved in the SPL. We will
explicitly check that for the decay paradimuonium ($\mu ^{+}\mu ^{-}$) into $%
\gamma e^{+}e^{-}$, correct SPL is restored through the interference of
oblique cuts with the standard approach vertical cuts amplitudes. This
analysis is done in the last section, along with the analogous decay $%
K_{S}\rightarrow \gamma e^{+}e^{-}$.

\subsubsection{Structure-dependent terms}

The violation of gauge invariance by bremsstrahlung-type amplitude is a very
well-known phenomenon in pion and kaon physics. One remembers that for a
decay process like $\pi ^{+}\left( K^{+}\right) \rightarrow e^{+}\nu \gamma $%
, the so-called Inner-Bremsstrahlung (IB) amplitudes combine with Structure
Dependent (SD) amplitudes to restore gauge invariance (see for example \cite
{NonCstF}, or \cite{SDChiral} for a modern Chiral Perturbation Theory point
of view, see also \cite{PDG} for measurements of structure dependent form
factors, and recent literature cited there).

What we here propose is a similar mechanism for positronium. On general
grounds, it was foreseeable that {\it some interplay between the bound state
dynamics and the decay process should appear, both ultimately involving
photons}. The present model allows us to give some constraints for the
general form of those structure terms, exactly like for pseudoscalar pion
decay. However, to obtain their exact forms is a very difficult problem.

To analyze the restoration of both the correct SPL behavior and gauge
invariance thanks to SD\ terms, we invoke Low's theorem in its original form
(see \cite{Low}, \cite{NonCstF}):

\begin{quote}
For a scalar decay process $A^{0}\rightarrow e^{+}\left( p_{1}\right)
e^{-}\left( p_{2}\right) \gamma \left( k\right) $ with both structure and
bremsstrahlung contributions, the complete amplitude behaves in the SPL as 
\begin{eqnarray}
{\cal M}^{\mu }\left( A^{0}\rightarrow e^{+}e^{-}\gamma \right)
&=&F_{Ae^{+}e^{-}}\left( p_{1},p_{2}\right) \left\{ \overline{u}\left(
p_{1}\right) \frac{2p_{1}^{\mu }+\gamma ^{\mu }%
\!\not\!%
k}{2p_{1}\cdot k}v\left( p_{2}\right) \right\}  \label{BremLow} \\
&&-F_{Ae^{+}e^{-}}\left( p_{1},p_{2}\right) \left\{ \overline{u}\left(
p_{1}\right) \frac{2p_{2}^{\mu }+%
\!\not\!%
k\gamma ^{\mu }}{2p_{2}\cdot k}v\left( p_{2}\right) \right\} +{\cal O}\left(
k^{0}\right)  \nonumber
\end{eqnarray}
if the Ward identity $k_{\mu }{\cal M}^{\mu }=0$ holds. Three important
consequences:

1-- There is no constant term, i.e. ${\cal O}\left( \left( k^{0}\right)
^{0}\right) $, arising from the form factor.

2-- The form factor appears as evaluated at the same point.

3-- When the process involves only neutral bosons, ${\cal M}\sim {\cal O}%
\left( k^{0}\right) $.
\end{quote}

To understand the content of the first point, let us expand $F_{B}\left(
p_{1}+l_{1},p_{2}\right) $ in (\ref{Brem}) around $l_{1}^{0}=0$. The
expansion of the form factor will bring a new constant term 
\begin{eqnarray*}
F_{B}\left( p_{1}+l_{1},p_{2}\right) \frac{2p_{1}^{\mu }+\gamma ^{\mu }%
\!\not\!%
l_{1}}{2p_{1}\cdot l_{1}} &\rightarrow &F_{B}\left( p_{1},p_{2}\right) \frac{%
2p_{1}^{\mu }+\gamma ^{\mu }%
\!\not\!%
l_{1}}{2p_{1}\cdot l_{1}} \\
&&+2ap_{1}^{\mu }\frac{\partial F_{B}\left( p_{1},p_{2}\right) }{\partial
p_{1}}+{\cal O}\left( l_{1}^{0}\right)
\end{eqnarray*}
for some $a$. It is this additional constant term that must be cancelled by
SD amplitudes to preserve gauge invariance, hence leading to Low's theorem (%
\ref{BremLow}).

The second point can be rephrased as follows: near $k^{0}=0$, the complete
amplitude behaves exactly as if the form factor was a constant (i.e. it is
evaluated at the symmetric point (\ref{SymPoint}) for all contributions).
This is the key. The result is neatly expressed as

\begin{quote}
{\it In the soft-photon limits, the loop amplitudes behave exactly like the
constant form factor loop model, because structure-dependent amplitudes
interfere with oblique cuts to cancel their dependences on the form factor
variations. As a result, positronium decay amplitudes vanish in the SPL,
since for a constant form factor loop model, oblique cuts interfere
destructively with vertical cuts.}
\end{quote}

This result is quite powerful, because the dynamics of each contribution is
manifest. This opens the way to an evaluation of their respective scalings
in $\alpha $ and $\gamma $, as we now discuss.

\subsubsection{Scaling and NRQED}

At first sight, all contributions (vertical, oblique and structure) should
be of comparable numerical size, all of them being of the same order in $%
\alpha $. This is in fact not at all the case. A very special feature of
electromagnetic (loosely) bound states is that the binding energy can be
expressed in terms of $\alpha $%
\[
\gamma \approx m\alpha /2 
\]
It is one of the great virtue of NRQED to provide a consistent framework to
re-order the perturbation series according to powers of $\alpha ^{n}\gamma
^{m}\rightarrow \alpha ^{n+m}$.

Another virtue of NRQED is that the starting point is the Bethe-Salpeter
loop model (\ref{BSdecay}), i.e. the same loop model as (\ref{OrthoLoop}).
This would mean that all the cuts should somehow be contained in NRQED
calculations. If it is true that lowest order NRQED starts by setting $%
q_{0}=0$ in (\ref{BSdecay}) (because kinetic energy is subleading compared
to momentum), systematic corrections to this approximation are then
calculated, and the four-dimensional loop is ''reconstructed''. In other
words, the lowest order NRQED result is our vertical cut results, while
oblique cuts correspond to higher order corrections.

Our approach could then be viewed as complementary to NRQED perturbative
expansions. Dispersion relations put forward basic principles, and should
provide interesting constraints on the systematics of NRQED.

Structure-dependent contributions are not contained in NRQED. Their form is
unknown and so is their numerical importance. Their existence means that the
issue of gauge invariance for the NRQED expansion is irrelevant, because
hopeless. This means that the ''reconstruction'' of the four-dimensional BS
loop is not constrained by gauge invariance. However, by showing that
structure-terms are contributing at the order $\alpha ^{3}$, one proves that
the spurious gauge dependence and violation of analyticity of NRQED is also
of that order, hence validating the present ${\cal O}\left( \alpha
^{2}\right) $ theoretical predictions for the rates.

Let us review the different contributions. What follows is not a rigorous
demonstration, but rather a qualitative analysis. Nevertheless, the exact
evaluation should not alter much our conclusions.

{\bf i- The vertical cuts} alone are sufficient to reproduce the known
lowest order decay rate \cite{OrePowell} 
\[
\Gamma \left( o\text{-}Ps\rightarrow \gamma \gamma \gamma \right) =\frac{%
2\left( \pi ^{2}-9\right) }{9\pi }\alpha ^{6}m 
\]
This result is obtained in the static limit, i.e. with $F_{B}\propto \delta
^{\left( 3\right) }\left( {\bf q}\right) $. Let us also give the
differential rate \cite{Adkins}, 
\begin{equation}
\frac{d\Gamma \left( o\text{-}Ps\rightarrow \gamma \gamma \gamma \right) }{%
dx_{1}}=\frac{2\alpha ^{6}m}{9\pi }\left[ \tfrac{2\left( 2-x_{1}\right) }{%
x_{1}}+\tfrac{2\left( 1-x_{1}\right) x_{1}}{\left( 2-x_{1}\right) ^{2}}%
+\left[ \tfrac{4\left( 1-x_{1}\right) }{x_{1}^{2}}-\tfrac{4\left(
1-x_{1}\right) ^{2}}{\left( 2-x_{1}\right) ^{3}}\right] \ln \left(
1-x_{1}\right) \right]  \label{ExactPS}
\end{equation}
with $x_{1}=l_{1}^{0}/2M$ the reduced photon energy. The spectrum (\ref
{ExactPS}) illustrates the violation of analyticity. Near zero, it vanishes
linearly 
\begin{equation}
\frac{d\Gamma \left( o\text{-}Ps\rightarrow \gamma \gamma \gamma \right) }{%
dx_{1}}\stackrel{x_{1}\sim 0}{\sim }\frac{5}{3}x_{1}  \label{Near0PS}
\end{equation}
while we would expect a $x_{1}^{3}$ behavior instead. Indeed, the phase
space alone is 
\begin{equation}
\frac{d\Gamma \left( o\text{-}Ps\rightarrow \gamma \gamma \gamma \right) }{dx%
}\sim \frac{2\alpha ^{6}m}{9\pi }\left[ 2x_{1}\right]  \label{ApproxPS}
\end{equation}
Since the amplitude should behave as $x_{1}$ near $x_{1}=0$, the
differential rate should exhibit a $x_{1}^{3}$ behavior in the SPL. As we
have seen, the additional contributions restore the correct SPL.

{\bf ii- The oblique cuts} are suppressed by three effects: they disappear
in the limit of vanishing binding energy ($\gamma \rightarrow 0$), the
dispersion relation integration range is reduced compared to the one for
vertical cuts, and the three-photon phase-space enhances vertical cuts
against oblique ones.

The first point is obvious: if the $e^{+}e^{-}$ pair emerges on-shell from
the bound state, there is no room for oblique cuts. In fact, this limit is
pathological, because oblique cuts are still there, but contribute only when
one of the photon energies is strictly zero, to enforce SPL. Anyway, their
complete contribution should somehow be proportional to the binding energy $%
\gamma $, i.e. the mass default of the $e^{+}e^{-}$ inside the positronium.

To explain the second point, consider the dispersion integral (\ref{DRob}).
The lower bound $s_{\min \left( l_{1}\right) }$ is increasing as $l_{1}^{0}$
increases. Indeed, if the photon is carrying away some energy, the initial
bound state mass squared $s$ must be greater to create on-shell intermediate 
$e^{+}e^{-}$. Therefore, oblique cuts should be decreasing functions of the
photon energies.

The first two effects combine to suppress oblique cut contributions
everywhere except near zero photon energy. Therefore, oblique cut
contributions to the decay rate will be further (substantially) suppressed
by the phase-space (\ref{ExactPS}), peaked for high photon energies.

Due to all those effects, one could state that oblique cuts should
contribute at the order $\gamma ^{2}\approx \alpha ^{2}$, i.e. ${\cal D}%
\approx \alpha ^{2}{\cal V}$. Since the argument is qualitative, it could
happen that they are enhanced for some reason (or further suppressed), but
we do not see any compelling evidence of such pathologies.

{\bf iii-} Finally, {\bf structure terms} can be estimated by taking the
conservative view that their importance is just what is required to preserve
gauge invariance and SPL. Since they compete with oblique cuts, they are of
the same magnitude, i.e. ${\cal O}\left( \alpha ^{2}\right) $. However,
structure terms account for the {\it variation} of the form factor. For a
Schr\"{o}dinger-like form factor 
\[
{\cal F}\left( {\bf q}^{2}\right) =\frac{8\pi \gamma }{\left( {\bf q}%
^{2}+\gamma ^{2}\right) ^{2}} 
\]
The variations will be a much more peaked function of ${\bf q}^{2}$, leading
to a smaller result for the dispersion integral (\ref{DRsd}). All in all, we
should gain a power of the binding energy ${\cal S}\approx \gamma {\cal D}$.
In terms of the fine-structure constant, structure term should be of order $%
\alpha ^{3}$, i.e. ${\cal S}\approx \alpha ^{3}{\cal V}$.

Of course, this is not very convincing. If the pion decay can be used as a
guide, structure terms should indeed be highly suppressed. However, the very
small mass gap between the parapositronium and orthopositronium could lead
to huge enhancements, if one imagines structure term dynamics as driven by
intermediate virtual parapositronium. \newline

In conclusion, we argue that, provided that no pathological effects come
into play, oblique cuts could be considered as small ${\cal O}\left( \alpha
^{2}\right) $ corrections, and structure dependent terms neglected at the
present theoretical precision level. Anyway, no answer to the
orthopositronium lifetime puzzle could be viewed as entirely satisfactory as
long as no definite estimation or calculation of structure-dependent
processes has been achieved.

\section{Form Factor and Parapositronium Decay}

We have established the correspondence between (\ref{Loop1}) and (\ref{TH3}%
). Let us construct an alternative, but equivalent, dispersion procedure
specific to the two-photon case that will be used in explicit calculations.

By computing the trace in (\ref{Loop1}), the tensor structure factorizes and
we are left with 
\begin{equation}
{\cal M}\left( p\text{-}Ps\rightarrow \gamma \gamma \right)
=8me^{2}\varepsilon ^{\mu \nu \rho \sigma }l_{1,\rho }l_{2,\sigma
}\varepsilon _{1\mu }^{*}\varepsilon _{2\nu }^{*}{\cal I}\left( M^{2}\right)
\label{AmpTwoPhotI}
\end{equation}
and the decay rate is expressed simply as 
\begin{equation}
\Gamma \left( p\text{-}Ps\rightarrow \gamma \gamma \right) =16\pi \alpha
^{2}m^{2}M^{3}\left| {\cal I}\left( M^{2}\right) \right| ^{2}
\label{WidthTwoPhot}
\end{equation}
In this equation, ${\cal I}\left( M^{2}\right) $ can be viewed as an
effective form factor, modeled as the electron-positron loop with the
coupling $F_{B}$. There is only one term in ${\cal I}\left( M^{2}\right) $
since the direct and crossed amplitudes are equal under $q\rightarrow -q,$
i.e. an allowed variable change as $F_{B}\left( q^{2},P\cdot q\right)
=F_{B}\left( q^{2},-P\cdot q\right) $, and we write 
\begin{equation}
{\cal I}\left( P^{2}\right) =\eta \int \frac{d^{4}q}{\left( 2\pi \right) ^{4}%
}F_{B}\frac{1}{\left( q-\frac{1}{2}P\right) ^{2}-m^{2}}\frac{1}{\left( q+%
\frac{1}{2}P\right) ^{2}-m^{2}}\frac{1}{\left( q-\frac{1}{2}P+l_{1}\right)
^{2}-m^{2}}  \label{TwoPhotInt}
\end{equation}
It is to evaluate the effective form factor ${\cal I}\left( P^{2}\right) $
that we will now use dispersion techniques. Remark that the factorization of
the tensor part is interesting, since gauge invariance is manifest, and that 
${\cal I}\left( P^{2}\right) $ is convergent while the amplitude (\ref{Loop1}%
) is superficially divergent.

The factor $\eta $ is introduced because there is a subtlety in the above
factorization. Indeed, there is an arbitrariness in the choice of variable
for the dispersion integral. This situation is well-known for the photon
vacuum polarization: 
\begin{equation}
\Pi ^{\mu \nu }\left( k^{2}\right) =\left( k^{2}g^{\mu \nu }-k^{\mu }k^{\nu
}\right) \Pi _{1}\left( k^{2}\right) =\left( g^{\mu \nu }-\frac{k^{\mu
}k^{\nu }}{k^{2}}\right) \Pi _{2}\left( k^{2}\right)  \label{PhotPol}
\end{equation}
where one writes a dispersion relation for $\Pi _{1}\left( k^{2}\right) $,
which is less divergent than $\Pi _{2}\left( k^{2}\right) $ due to the
factorization of the tensor structure. The analogue of (\ref{PhotPol}) for
our case is 
\[
{\cal M}\left( p\text{-}Ps\rightarrow \gamma \gamma \right)
=8me^{2}\varepsilon ^{\mu \nu \rho \sigma }l_{1,\rho }l_{2,\sigma
}\varepsilon _{1\mu }^{*}\varepsilon _{2\nu }^{*}{\cal I}_{1}\left(
M^{2}\right) =8me^{2}\varepsilon ^{\mu \nu \rho \sigma }\frac{l_{1,\rho }}{M}%
\frac{l_{2,\sigma }}{M}\varepsilon _{1\mu }^{*}\varepsilon _{2\nu }^{*}{\cal %
I}_{2}\left( M^{2}\right) 
\]
We will choose to write a dispersion relation for ${\cal I}_{2}\left(
M^{2}\right) $, and this corresponds to the choice $\eta =P^{2}/M^{2}$. This
seemingly arbitrary choice is in fact necessary to be consistent with the
general development, see (\ref{TH1}), (\ref{TH2}). In other words, the
dispersion relation was built on the whole amplitude, so it is clear that
the photon momenta appearing in the tensor structure were also incorporated,
i.e. they were reduced as $l_{1,\rho }\rightarrow \overline{l_{1,\rho }}%
\times \sqrt{P^{2}}/2$. That is the reason why we must include a factor $%
P^{2}$ into the effective form factor ${\cal I}\left( P^{2}\right) $.

Let us give a general expression for ${\cal I}\left( P^{2}\right) $ as a
dispersion integral. The loop integral ${\cal I}\left( P^{2}\right) $ has an
imaginary part obtained by cutting the propagators 
\[
\func{Im}{\cal I}\left( P^{2}\right) =\frac{1}{2}\frac{P^{2}}{M^{2}}\int 
\frac{d^{4}q}{\left( 2\pi \right) ^{4}}F_{B}\frac{2\pi i\delta \left( \left(
q-\frac{1}{2}P\right) ^{2}-m^{2}\right) 2\pi i\delta \left( \left( q+\frac{1%
}{2}P\right) ^{2}-m^{2}\right) }{\left( q-\frac{1}{2}P+l_{1}\right)
^{2}-m^{2}} 
\]
Proceeding exactly like in the general case (see appendix), using an
unsubstracted DR, we reach the two equivalent forms 
\begin{mathletters}
\begin{eqnarray}
{\cal I}\left( M^{2}\right) &=&\frac{C\phi _{o}}{2M^{2}}\frac{1}{32\pi ^{2}}%
\int_{4m^{2}}^{+\infty }ds{\cal F}\left( s/4-m^{2}\right) \ln \left[ \frac{1+%
\sqrt{1-\frac{4m^{2}}{s}}}{1-\sqrt{1-\frac{4m^{2}}{s}}}\right]
\label{ThirdDRa} \\
&=&\frac{C\phi _{o}}{2M^{2}}\int \frac{d^{3}{\bf q}}{\left( 2\pi \right) ^{3}%
}{\cal F}\left( {\bf q}^{2}\right) \frac{1}{\sqrt{\left| {\bf q}\right|
^{2}+m^{2}}+\left| {\bf q}\right| \cos \theta _{{\bf q}}}  \label{ThirdDRb}
\end{eqnarray}
where we have set $F_{B}\equiv C\phi _{o}{\cal F}\left( {\bf q}^{2}\right)
\left( {\bf q}^{2}+\gamma ^{2}\right) $ (see \ref{Loop3}). The first
expression is obtained by integration over the angular variable $\theta _{%
{\bf q}}$.

This is our third representation for the same decay amplitude : the first is
the loop integral (\ref{Loop1}), the second is the well-known amplitude (\ref
{TH3}) (or (\ref{STD2}), with no approximation for projectors) viewed as a
dispersion integral for the amplitude, and the third is the present
dispersion integral (\ref{ThirdDRa},b) for the effective loop form factor $%
{\cal I}\left( M^{2}\right) $. All three procedures are strictly equivalent
to each other.

\subsection{Decay Rate in the Static Limit}

Let us first analyze the static limit, i.e. $\gamma ^{2}\rightarrow 0$ for
the form factor. To compute the decay rate in that limit, we do not need to
specify $F_{B}$ yet. We just need to know that the function ${\cal F}\left( 
{\bf k}^{2}\right) $ is normalized to unity and behaves as a delta function
of the momentum in the limit of vanishing binding energy: 
\end{mathletters}
\begin{equation}
\int \frac{d^{3}{\bf k}}{\left( 2\pi \right) ^{3}}{\cal F}\left( {\bf k}%
^{2}\right) =1,\quad \,\,\stackunder{\gamma \rightarrow 0}{\lim }{\cal F}%
\left( {\bf k}^{2}\right) =\left( 2\pi \right) ^{3}\delta ^{\left( 3\right)
}\left( {\bf k}\right)  \label{DeltaLimit}
\end{equation}
By setting ${\cal F}\left( {\bf q}^{2}\right) =\left( 2\pi \right)
^{3}\delta ^{\left( 3\right) }\left( {\bf q}\right) $ in (\ref{TwoPhotInt}),
we must recover exactly the lowest order decay rate 
\begin{equation}
\Gamma \left( p\text{-}Ps\rightarrow \gamma \gamma \right) =\frac{1}{2}%
\alpha ^{5}m  \label{LowestPara}
\end{equation}
In some sense, the static limit serves as a boundary condition for the
behavior of ${\cal F}$.

Using (\ref{ThirdDRb}), we get 
\[
{\cal I}\left( M^{2}\right) =\frac{C\phi _{o}}{2M^{2}}\int \frac{d^{3}{\bf q}%
}{\left( 2\pi \right) ^{3}}\left( 2\pi \right) ^{3}\delta ^{\left( 3\right)
}\left( {\bf q}\right) \frac{1}{\sqrt{\left| {\bf q}\right| ^{2}+m^{2}}%
+\left| {\bf q}\right| \cos \theta }=\frac{C\phi _{o}}{2M^{2}}\left[ \frac{1%
}{m}\right] 
\]
Importantly, this result is independent of the binding energy : the loop do
not introduce any corrections in the static limit. The decay rate in that
limit is therefore: 
\begin{equation}
\Gamma \left( p\text{-}Ps\rightarrow \gamma \gamma \right) =\frac{1}{2}%
\alpha ^{5}m\left( \frac{m^{2}}{M}C^{2}\right)  \label{WidthCMatch}
\end{equation}
with $\left| \phi _{0}\right| ^{2}=\alpha ^{3}m^{3}/8\pi $. It remains to
match $C$ such that purely kinematic corrections vanish ($M$ factors in the
above formula arise from products like $\left( l_{1}\cdot l_{2}\right)
=M^{2}/2$ and from the $1/2M$ decay width factor, while $m$ comes from
electron propagators in the loop and from the wavefunction $\phi _{0}$).
With the definition $C=\sqrt{M}/m$, the decay rate is exactly $\frac{1}{2}%
\alpha ^{5}m$ as it should. In other words, the value for $C$ obtained by
matching (\ref{TH3}) and (\ref{STD2}) is such that no correction arises from
factors $M/m$ in the static limit.

To conclude, let us repeat that we have not specified the form factor. This
means that any form factor which has a three-dimensional delta function
limit for $\gamma ^{2}\rightarrow 0$ gives the correct lowest order decay
rate $\frac{1}{2}\alpha ^{5}m$. In the following, we shall present three
forms, all built on the Schr\"{o}dinger momentum wavefunction.

\subsection{Schr\"{o}dinger Form Factors and Binding Energy Corrections}

The formula (\ref{ThirdDRa}) is a very simple expression: the decay rate (%
\ref{WidthTwoPhot}) for any form factor is obtained by a one-dimensional
integral. This is a very nice calculational tool that we will now use. We
will go through different calculations of $\Gamma \left( p\text{-}%
Ps\rightarrow \gamma \gamma \right) $, obtained for specific choices of $%
F_{B}$ (or equivalently, ${\cal F}\left( {\bf q}^{2}\right) $). Namely: 
\begin{mathletters}
\begin{eqnarray}
{\cal F}_{I}\left( {\bf q}^{2}\right) &=&\dfrac{8\pi \gamma }{\left( {\bf q}%
^{2}+\gamma ^{2}\right) ^{2}}  \label{Schro1} \\
{\cal F}_{II}\left( {\bf q}^{2}\right) &=&\dfrac{32\pi \gamma ^{3}}{\left( 
{\bf q}^{2}+\gamma ^{2}\right) ^{3}}  \label{Schro2} \\
{\cal F}_{III}\left( {\bf q}^{2}\right) &=&\frac{2\gamma }{\left| {\bf q}%
\right| }\arctan \frac{\left| {\bf q}\right| }{\gamma }\;\times \;{\cal F}%
_{I}\left( {\bf q}^{2}\right)  \label{Schro3}
\end{eqnarray}
where $\gamma ^{2}=m^{2}-M^{2}/4\approx m^{2}\alpha ^{2}/4$ related to the
binding energy $E_{B}=M-2m=-m\alpha ^{2}/4$. All these form factors satisfy
the delta limit property (\ref{DeltaLimit}), but are differently peaked near 
$\left| {\bf q}\right| =0$ (see figure 7). As a result, we expect that all
of them will reproduce the lowest order decay rate (\ref{LowestPara}), but
will introduce some $\gamma $-dependent corrections proportional to their
spreading around $\left| {\bf q}\right| =0$. Such corrections can then be
expressed in terms of the fine-structure constant, yielding a corrected
lowest order rate of 
\end{mathletters}
\[
\Gamma \left( p\text{-}Ps\rightarrow \gamma \gamma \right) =\frac{1}{2}%
\alpha ^{5}m\left( 1+\delta _{\Gamma }\left( {\cal F}\right) \right) 
\]

The physics of each form factor is discussed below. For now, let us use the
formula (\ref{ThirdDRa}) to get the effective form factors and decay rates
corrections $\delta _{\Gamma }\left( {\cal F}_{I,II,III}\right) $. For the
first one, we find 
\begin{equation}
{\cal I}_{I}\left( M^{2}\right) =\frac{C\phi _{o}}{M^{3}}\frac{2}{\pi }%
\arctan \frac{M}{2\gamma }  \label{ReIntSchro1}
\end{equation}
The integral needed for this calculation is (see \cite{TextBook2}) 
\begin{equation}
\int_{0}^{1}\frac{dx}{x_{o}-x}\,\,\ln \left[ \frac{1+\sqrt{1-x}}{1-\sqrt{1-x}%
}\right] \stackrel{x_{o}>1}{=}2\arctan ^{2}\frac{1}{\sqrt{x_{o}-1}}
\label{DispDJ}
\end{equation}
and its derivative $\partial /\partial x_{o}$. The decay rate is 
\begin{equation}
\Gamma _{I}\left( p\text{-}Ps\rightarrow \gamma \gamma \right) =\frac{1}{2}%
\alpha ^{5}m\left( \frac{4m^{2}}{M^{2}}\right) \left( \frac{2}{\pi }\arctan 
\frac{M}{2\gamma }\right) ^{2}  \label{CorrRate}
\end{equation}
where we used $\left| \phi _{0}\right| ^{2}=\alpha ^{3}m^{3}/8\pi $ and $C=%
\sqrt{M}/m$. By expanding this result around $\gamma =0$, and expressing
corrections as a series in $\alpha $, we recover the standard result as
zeroth order 
\begin{eqnarray*}
\Gamma _{I}\left( p\text{-}Ps\rightarrow \gamma \gamma \right) &=&\frac{1}{2}%
\alpha ^{5}m\left( 1-\frac{\alpha }{\pi }+\frac{1}{8}\alpha ^{2}+{\cal O}%
\left( \alpha ^{3}\right) \right) ^{2} \\
&\approx &\frac{1}{2}\alpha ^{5}m\left( 1-0.637\alpha +0.351\alpha ^{2}+%
{\cal O}\left( \alpha ^{3}\right) \right)
\end{eqnarray*}
Proceeding similarly for the other two, using (\ref{DispDJ}) for ${\cal F}%
_{II}$, and numerical integration for ${\cal F}_{III}$, we find 
\begin{eqnarray*}
\Gamma _{II}\left( p\text{-}Ps\rightarrow \gamma \gamma \right) &\approx &%
\frac{1}{2}\alpha ^{5}m\left( 1-0.25\alpha ^{2}\right) \\
\Gamma _{III}\left( p\text{-}Ps\rightarrow \gamma \gamma \right) &\approx &%
\frac{1}{2}\alpha ^{5}m\left( 1-1.73\alpha ^{2}\right)
\end{eqnarray*}
If we naively combine the present corrections with radiative corrections up
to order $\alpha ^{2}\ln \alpha $: 
\[
\Gamma _{p\text{-}Ps}=\frac{\alpha ^{5}m}{2}\left( 1-\left( 5-\frac{\pi ^{2}%
}{4}\right) \frac{\alpha }{\pi }+2\alpha ^{2}\ln \frac{1}{\alpha }+\delta
_{\Gamma }\left( {\cal F}\right) \right) 
\]
the theoretical value is modified as 
\[
\begin{tabular}{l|l|l}
Lowest-order & Corrections & Decay Rate, with \\ 
Form factor & $\delta _{\Gamma }$ & Radiative Corrections and $\delta
_{\Gamma }$ \\ \hline
$Static\;Limit$ & $0$ & $7.9895\times 10^{9}\sec ^{-1}$ \\ 
${\cal F}_{I}$ & $-0.637\alpha $ & $7.9527\times 10^{9}\sec ^{-1}$ \\ 
${\cal F}_{II}$ & $-0.25\alpha ^{2}$ & $7.9894\times 10^{9}\sec ^{-1}$ \\ 
${\cal F}_{III}$ & $-1.73\alpha ^{2}$ & $7.9887\times 10^{9}\sec ^{-1}$ \\ 
\hline
Experiments &  & $\left( 7.9909\pm 0.0017\right) \times 10^{9}\sec ^{-1}$%
\end{tabular}
\]
where the experimental measurement (\ref{ExppPs}) is also included for
comparison.

\subsubsection{Discussion}

In view of the above table, it appears that the first form factor ${\cal F}%
_{I}$ leads to a too small value for $\Gamma \left( p\text{-}Ps\rightarrow
\gamma \gamma \right) $, since it introduces new order $\alpha $
corrections. The problem is in the form factor, which does not converge fast
enough towards the static limit delta. In other words, for a given $\gamma
^{2}$, ${\cal F}_{I}\left( {\bf q}^{2}\right) $ is not enough peaked around $%
\left| {\bf q}\right| =0$. On the other hand, the more peaked ${\cal F}_{II}$
and ${\cal F}_{III}$ form factors lead to acceptable corrections.

It is time now to consider the physical content of each form factor. In
summary, the ${\cal F}_{I}$ is just the Shr\"{o}dinger momentum wavefunction
for the bound state, the ${\cal F}_{II}$ form factor has no clear
signification and only serves the purpose of illustration, while ${\cal F}%
_{III}$ can be viewed as a Coulomb binding corrected Schr\"{o}dinger form
factor, in the spirit of the Sommerfeld factor. Of course, one could argue
that the form factor to be used must be linked somehow to the Bethe-Salpeter
wavefunction. In fact, it is the first form factor ${\cal F}_{I}$ that
emerges from the reduction of the Barbieri-Remiddi wavefunction through
dispersion relations (\ref{TH3}). Despite of that, we will argue that it is
the ${\cal F}_{III}$ form factor that must be considered if one is
interested in ${\cal O}\left( \alpha \right) $ accuracy (this discussion is
much inspired from \cite{Adkins}).

The fact that the exact integration of the ${\cal F}_{I}$ form factor leads
to excessive order $\alpha $ corrections $\delta _{\Gamma }\left( {\cal F}%
_{I}\right) $ is not new. Let us stress first that those corrections are $%
\gamma $-dependent (see (\ref{CorrRate})), i.e. {\it it is a binding energy
effect}. Such effects, being non-perturbative, are quite difficult to be
dealt with. Even in the ${\cal O}\left( \alpha \right) $ calculation of \cite
{Adkins}, much space is devoted to the discussion of such effects.
Double-countings plague bound state calculations, and it is quite possible
that the present excessive ${\cal O}\left( \alpha \right) $ corrections
should be discarded because they are already taken into account in ${\cal O}%
\left( \alpha \right) $ calculations \cite{LargeRelCorr}.

Indeed, the standard approach to these excessive corrections is simply to
discard them. Either they are assumed to be part of higher order radiative
corrections to the scattering process (\cite{Tomozawa}, \cite{Adkins}), or
alternatively, they are considered as part of the relativistic corrections
to the bound state wavefunction \cite{LargeRelCorr}, \cite{LargeRelCorr2}.
In any case, they are interpreted as accounting for some modification of the
intermediate $e^{+}e^{-}$ state: either as emitted from the bound state,
either as entering the scattering process.

This dual view is further enhanced by the following: those additional
corrections disappear in the static limit, i.e. when $\gamma ^{2}\rightarrow
0$, so they appear as originating from the bound state dynamics.
Alternatively, looking back at (\ref{ThirdDRb}), it appears that the
electron propagator between the two photons plays a crucial role in
introducing the additional corrections. If it is frozen (i.e. replaced by $%
1/m$), the corrections disappear. In other words, one could view the decay
process instead of the bound state dynamics as responsible for the excessive
corrections.

From the point of view of the BS equation, things are quite clear. The
Schr\"{o}dinger momentum wavefunction is built as the Coulomb photon
exchange ladder approximation (see figure 8a). Relativistic corrections to
the wavefunction must be considered to account for the covariant photon
exchange. This corresponds to the relativistic corrections point of view.
The radiative corrections view arises by considering the binding graph
(figure 8b). As shown for example in \cite{Adkins}, this ${\cal O}\left(
\alpha \right) $ radiative correction for the scattering amplitude
reproduces the lowest order decay amplitude, because its Coulomb photon
exchange part has already been accounted for in the construction of the
wavefunction.

What we propose is to view the third form factor as a representation of the
modification of the intermediate $e^{+}e^{-}$ state. When dispersion
relations are used, one consider intermediate states as asymptotic states,
and integrate over the corresponding phase-space. This is the application of
the optical theorem for absorptive parts. Let us assume that, due to the
modification of the asymptotic Hilbert space by the long-range Coulomb
interactions, a correction factor must be introduced in 
\begin{equation}
{\cal M}\left( p\text{-}Ps\rightarrow \gamma \gamma \right) \sim \int d^{3}%
{\bf q\;}\psi \left( {\bf q}^{2}\right) \times S\left( {\bf q}\right) \times 
{\cal M}\left( e_{{\bf q}}^{-}e_{-{\bf q}}^{+}\rightarrow \gamma \gamma
\right)  \label{PhaseSpaceMod}
\end{equation}
and that this factor is 
\begin{equation}
S\left( {\bf q}\right) =\frac{2\gamma }{\left| {\bf q}\right| }\arctan \frac{%
\left| {\bf q}\right| }{\gamma }  \label{CorrS}
\end{equation}
In some sense, one could think of $S\left( {\bf q}\right) $ as the
Sommerfeld factor of Harris and Brown \cite{Harris}. Due to long range
Coulomb interactions, the wavefunction at contact $\phi _{o}$ is
''renormalized'' by 
\[
\left| \phi _{o}\right| ^{2}\rightarrow \left| \phi _{o}\right| ^{2}\frac{%
2\pi \alpha /v}{1-e^{-2\pi \alpha /v}}\approx \left| \phi _{o}\right|
^{2}\left( 1+\frac{\pi \alpha }{v}\right) 
\]
by using $v>>\alpha $, i.e. $\left| {\bf q}\right| >>\gamma $, with $v$ the
center-of-mass velocity. Correspondingly, $S\left( {\bf q}\right) $ for
small binding energy is 
\[
S\left( {\bf q}\right) \approx \frac{\pi \gamma }{\left| {\bf q}\right| }%
\approx \frac{\pi \alpha }{2v} 
\]
The factor of $2$ arises because $S\left( {\bf q}\right) $ is defined at the
amplitude level, while the Sommerfeld factor is at the decay rate level.

The same arctangent factor arises in the work of \cite{Tomozawa}, \cite
{Adkins} on ${\cal O}\left( \alpha \right) $ radiative corrections, by a
detailed analysis of the binding graph and its Coulombic part (which is,
after all, an initial state Coulombic interaction).

In conclusion, if one is interested in the lowest approximation (i.e. the $%
m\alpha ^{5}/2$) the static limit is just fine. If one wants the lowest
approximation to ${\cal O}\left( \alpha \right) $ accuracy, one must
consider it as arising from the ${\cal O}\left( \alpha \right) $ binding
graph, as demonstrated by Adkins \cite{Adkins}. In other words, by taking 
{\it only} ${\cal O}\left( \alpha \right) $ diagrams for the scattering
amplitude $e^{+}e^{-}\rightarrow \gamma \gamma $, one finds 
\[
\Gamma \left( p\text{-}Ps\rightarrow \gamma \gamma \right) =\frac{1}{2}%
m\alpha ^{5}\left( 1-\frac{\alpha }{\pi }\left( 5-\frac{\pi ^{2}}{4}\right)
\right) +{\cal O}\left( \alpha ^{2}\right) 
\]
The $1$ in the above expression is the lowest approximation to ${\cal O}%
\left( \alpha \right) $ accuracy, arising from the ${\cal O}\left( \alpha
\right) $ amplitude. To reproduce it using our effective form factor method,
the Coulomb corrected Schr\"{o}dinger form factor ${\cal F}_{III}$ is to be
used, as in \cite{Adkins}.

What we gained using our method is a better understanding of its link with
the Sommerfeld factor, i.e. asymptotic Coulomb interactions. ${\cal F}_{III}$
being more peaked than ${\cal F}_{I}$, the suppression of the intermediate
phase-space is interpreted as a manifestation of the iterative structure of
Bethe-Salpeter construction of the wavefunction.

Therefore, the method of \cite{Adkins} is nothing but the rephrasing of that
of Harris and Brown \cite{Harris}. However, the method of \cite{Adkins} or
the present integration of ${\cal F}_{III}$ is surely more appropriate,
because one does not rely on static limits, hence avoids the well-known
static divergence $1/v$ of radiative corrections to $e^{-}e^{+}\rightarrow
\gamma \gamma $ \cite{RadCorr}.

\subsubsection{Improving the perturbation series}

It is interesting to reverse the previous argument. Instead of curing the
form factor so that one avoids double-counting, it would be more adequate to
modify radiative corrections, keeping the ${\cal F}_{I}$ form factor.
Indeed, the form factor accounts for most of the ${\cal O}\left( \alpha
\right) $ corrections. We found 
\[
\Gamma _{I}\left( p\text{-}Ps\rightarrow \gamma \gamma \right) =\frac{\alpha
^{5}m}{2}\left( 1-2\frac{\alpha }{\pi }+{\cal O}\left( \alpha ^{2}\right)
\right) \approx \frac{\alpha ^{5}m}{2}\left( 1-0.637\alpha +{\cal O}\left(
\alpha ^{2}\right) \right) 
\]
while complete corrections are (\ref{Theor}) 
\[
\Gamma _{p\text{-}Ps}=\frac{\alpha ^{5}m}{2}\left( 1-\left( 5-\frac{\pi ^{2}%
}{4}\right) \frac{\alpha }{\pi }+{\cal O}\left( \alpha ^{2}\right) \right)
\approx \frac{\alpha ^{5}m}{2}\left( 1-0.806\alpha +{\cal O}\left( \alpha
^{2}\right) \right) 
\]
In other words, the decay rate can be expressed as 
\begin{eqnarray}
\Gamma _{p\text{-}Ps} &=&\frac{\alpha ^{5}m}{2}\frac{4m^{2}}{M^{2}}\left( 
\frac{2}{\pi }\arctan \frac{M}{2\gamma }\right) ^{2}\left( 1-\left( 3-\frac{%
\pi ^{2}}{4}\right) \frac{\alpha }{\pi }+{\cal O}\left( \alpha ^{2}\right)
\right)   \nonumber \\
&\approx &\frac{\alpha ^{5}m}{2}\frac{4m^{2}}{M^{2}}\left( \frac{2}{\pi }%
\arctan \frac{M}{2\gamma }\right) ^{2}\left( 1-0.170\alpha +{\cal O}\left(
\alpha ^{2}\right) \right)   \label{Improved}
\end{eqnarray}
This last form is very interesting because binding energy corrections (i.e. $%
\gamma $-dependent) are singled out, while the resulting purely radiative
corrections are much reduced. This means that one could, at least in
principle, express the decay rate as non-perturbative binding energy
corrections times a rapidly converging perturbation series of radiative
corrections. This is exactly what we have achieved to order $\alpha $. 

It is worthwhile to note that (\ref{Improved}) also accounts for most of the 
${\cal O}\left( \alpha ^{2}\right) $ corrections. Omitting logarithmic
corrections, the decay rate is given by 
\begin{eqnarray}
\Gamma _{p\text{-}Ps} &\approx &\frac{\alpha ^{5}m}{2}\left( 1-0.806\alpha
+0.52\alpha ^{2}+{\cal O}\left( \alpha ^{3}\right) \right)   \nonumber \\
&\approx &\frac{\alpha ^{5}m}{2}\frac{4m^{2}}{M^{2}}\left( \frac{2}{\pi }%
\arctan \frac{M}{2\gamma }\right) ^{2}\left( 1-0.170\alpha +0.06\alpha ^{2}+%
{\cal O}\left( \alpha ^{3}\right) \right)   \label{Improved2}
\end{eqnarray}
This appears as coincidental (especially in view of the many interferences
occurring among ${\cal O}\left( \alpha ^{2}\right) $ corrections, see \cite
{RecentCorr}), but it is nevertheless intriguing. It may be the signal that
an improved basis can be built for the positronium perturbative calculation,
if one avoids NRQED systematic $\alpha $ expansion of binding energy
effects. (Remark that this formulation does not cause any problem regarding
logarithmic corrections; the modification amount only to a reduction of the
coefficient $C_{p}$ of the $\alpha ^{3}\ln \alpha $ term to $-3.919\left(
1\right) $ from $-7.919\left( 1\right) $)

Of course, it would be very interesting to build the corresponding
expression for orthopositronium. Unfortunately, as extensively discussed,
both oblique cuts and (unknown) structure-dependent contributions render the
calculation highly challenging.

\section{Paradimuonium Decay into $\gamma e^{+}e^{-}$}

The present section concerns paradimuonium, the singlet $\mu ^{+}\mu ^{-}$
bound state \cite{Dimuonium}. This state has not been observed yet. The
reason why we consider the decay $p$-$Dm\rightarrow \gamma e^{+}e^{-}$ is
that it is the simplest process where a photon is kinematically allowed to
have a vanishing energy (Note that the positronium decay $p$-$Ps\rightarrow
\gamma \nu \overline{\nu }$ via a $Z^{0}$ has the same dynamics, but has not
been observed either). Therefore, we will be able to illustrate that the
loop model has a correct soft-photon behavior in the case of parapositronium
or paradimuonium decay.

The decay $p$-$Dm\rightarrow e^{+}e^{-}\gamma $ is shown in figure 9. The
simplicity of the process comes from the pseudoscalar nature of the
paradimuonium, which allows a manifestly gauge invariant treatment
throughout (hence the absence of structure-dependent terms). The amplitude
for $p$-$Dm\rightarrow e^{+}e^{-}\gamma $ can be directly obtained from the
positronium to two photon decay amplitude (\ref{AmpTwoPhotI}), (\ref
{TwoPhotInt}):

\begin{equation}
{\cal M}\left( p\text{-}Dm\rightarrow e^{+}e^{-}\gamma \right)
=8me^{3}\varepsilon ^{\mu \nu \rho \sigma }k_{\rho }t_{\sigma }\varepsilon
_{\mu }^{*}\left( k\right) \frac{\left\{ \overline{u}\left( p\right) \gamma
_{\nu }v\left( p^{\prime }\right) \right\} }{t^{2}+i\varepsilon }{\cal I}%
\left( M^{2},x\right)  \label{TensorFactor}
\end{equation}
with $x=2k^{0}/M$ the reduced photon energy, $M$ the dimuonium mass and $m$
the muon mass.\ The effective form factor loop integral is given by 
\[
{\cal I}\left( P^{2},x\right) =\eta \int \frac{d^{4}q}{\left( 2\pi \right)
^{4}}F_{B}\frac{1}{\left( q-\frac{1}{2}P\right) ^{2}-m^{2}}\frac{1}{\left( q+%
\frac{1}{2}P\right) ^{2}-m^{2}}\frac{1}{\left( q-\frac{1}{2}P+k\right)
^{2}-m^{2}} 
\]
where $\eta =P^{2}/M^{2}$ and $t^{2}=P^{2}(1-x)$ (the only difference with $%
p $-$Ps\rightarrow \gamma \gamma $ being that $t^{2}\neq 0$). Gauge
invariance is ensured by the factorized tensor structure, i.e. by the
antisymmetric Levi-Civita tensor.

From the decay amplitude, the differential width is obtained as 
\begin{equation}
\frac{d\Gamma \left( p\text{-}Dm\rightarrow e^{+}e^{-}\gamma \right) }{dx}=%
\frac{16\alpha ^{3}}{3}m^{2}\,M^{3}\left| {\cal I}\left( M^{2},x\right)
\right| ^{2}\rho \left( x,a\right)   \label{DiffRatepdm}
\end{equation}
with the phase space spectrum 
\begin{equation}
\rho \left( x,a\right) =\sqrt{1-\frac{a}{1-x}}\left[ a+2\left( 1-x\right)
\right] \frac{x^{3}}{\left( 1-x\right) ^{2}}  \label{PhaseSpace}
\end{equation}
where $a=4m_{e}^{2}/M^{2}$, and the bounds on $x$ are $\left[ 0,1-a\right] $%
. This spectrum is highly peaked just below $1-a$, as shown on figure 10.

We are going to calculate the integral ${\cal I}\left( M^{2},x\right) $
using dispersion relations. Because $t^{2}\neq 0$ (i.e. $x\neq 1$), two
different types of cuts will contribute (figure 11). We have shown that
considering the vertical cuts is strictly equivalent to the decay amplitude
calculation done using formula (\ref{TH3}) where $\Gamma $ is now the
(amputated) scattering amplitude $\mu ^{+}\left( k^{\prime }\right) \mu
^{-}\left( k\right) \rightarrow \gamma \gamma ^{*}\rightarrow
e^{+}e^{-}\gamma $ with on-shell muons.

What we are going to show is that {\it the amplitude for }$p${\it -}$%
Dm\rightarrow e^{+}e^{-}\gamma ${\it \ has a correct soft-photon limit only
if we take all the cuts into account}. By this we mean that each cut gives a
contribution to the amplitude that behaves as a constant when the photon
energy goes to zero. The combination of the vertical and oblique cuts, on
the contrary, forces the amplitude to vanish in that limit. We thus recover
the analytical behaviour expected from Low's theorem \cite{Low} for the
decay $p$-$Dm\rightarrow \gamma \gamma ^{*}$ involving only neutral bosons.

\subsection{Dispersion Relations and Soft photon limit}

In this section, we will analyze properties of the effective form factor $%
{\cal I}\left( P^{2},x\right) $. To keep the discussion as general as
possible, we will not specify the form factor $F_{B}$.

{\bf i- The absorptive part} is found by cutting the relevant propagators as 
\begin{eqnarray*}
\func{Im}{\cal I}_{1}\left( P^{2},x\right) &=&\eta \int \frac{d^{4}q}{%
2\left( 2\pi \right) ^{4}}F_{B}\frac{2\pi i\delta \left( \left( q-\frac{1}{2}%
P\right) ^{2}-m^{2}\right) 2\pi i\delta \left( \left( q+\frac{1}{2}P\right)
^{2}-m^{2}\right) }{\left( q-\frac{1}{2}P+k\right) ^{2}-m^{2}} \\
\func{Im}{\cal I}_{2}\left( P^{2},x\right) &=&\eta \int \frac{d^{4}q}{%
2\left( 2\pi \right) ^{4}}F_{B}\frac{2\pi i\delta \left( \left( q+\frac{1}{2}%
P\right) ^{2}-m^{2}\right) 2\pi i\delta \left( \left( q-\frac{1}{2}%
P+k\right) ^{2}-m^{2}\right) }{\left( q-\frac{1}{2}P\right) ^{2}-m^{2}}
\end{eqnarray*}
for the vertical and oblique cuts, respectively. By a straightforward
integration, the first expression gives ($s=P^{2}$): 
\begin{equation}
\func{Im}{\cal I}_{1}\left( s,x\right) =\frac{\eta }{s}\frac{1}{16\pi x}%
F_{B}\left( q_{0}=0,\left| {\bf q}\right| =\sqrt{s/4-m^{2}}\right) \ln
\left[ \frac{1+\sqrt{1-4m^{2}/s}}{1-\sqrt{1-4m^{2}/s}}\right] \theta \left(
s-4m^{2}\right)  \label{Cut1Imaginary}
\end{equation}
while the second one cannot be completely integrated without specifying $%
F_{B}$: 
\begin{equation}
\func{Im}{\cal I}_{2}\left( s,x\right) =\frac{\eta }{s}\frac{1}{16\pi x}%
\int_{q_{\min }}^{q_{\max }}\frac{dq_{0}}{q_{0}}F_{B}\left( q_{0},\left| 
{\bf q}\right| =\sqrt{q_{0}^{2}+q_{0}\sqrt{s}+s/4-m^{2}}\right) \theta
\left( s-\frac{4m^{2}}{1-x}\right)  \label{Cut2Imaginary}
\end{equation}
with the bounds given by 
\begin{equation}
q_{\min }=-\tfrac{x\sqrt{s}}{4}\left( 1+\sqrt{1-\tfrac{4m^{2}}{s\left(
1-x\right) }}\right) ,\;q_{\max }=-\tfrac{x\sqrt{s}}{4}\left( 1-\sqrt{1-%
\tfrac{4m^{2}}{s\left( 1-x\right) }}\right)  \label{DispEEGBounds}
\end{equation}

Interestingly, the second cuts contribute for $q_{0}\neq 0$. This is in
sharp contrast with the two-photon decay, since there only the first cuts
exist. Further, approximated bound state wavefunctions where a $\delta
\left( q_{0}\right) $ appears cannot be used (see \cite{Adkins} to \cite{BS2}%
), and one should revert to the full four-dimensional Bethe-Salpeter
wavefunction (for example the Barbieri-Remiddi one \cite{BarbRem}).

{\bf ii- In the soft photon limit}, the combination $\func{Im}{\cal I}%
_{1}\left( s,x\right) +\func{Im}{\cal I}_{2}\left( s,x\right) $ behaves as a
constant when $x\rightarrow 0$ despite the fact that each cut diverges in
that limit (see the $1/x$ in (\ref{Cut1Imaginary}) and (\ref{Cut2Imaginary}%
)\thinspace ). This will guarantee that the imaginary part of the whole
amplitude vanishes in the soft photon limit thanks to the presence of $%
k_{\rho }$ in the tensor structure of (\ref{TensorFactor}). The following
simple formula gives the behaviour of the imaginary part when $x\rightarrow 0
$%
\[
\func{Im}{\cal I}_{1}\left( s,x\right) +\func{Im}{\cal I}_{2}\left(
s,x\right) \stackrel{x\rightarrow 0}{=}\frac{\eta }{s}\frac{1}{16\pi }\left[ 
\frac{f_{B}}{\sqrt{1-\frac{4m^{2}}{s}}}+\frac{\sqrt{s-4m^{2}}}{2}%
f_{B}^{\prime }\right] +{\cal O}\left( x\right) 
\]
where 
\begin{eqnarray*}
f_{B} &=&F_{B}\left( 0,{\bf q}^{2}=s/4-m^{2}\right)  \\
f_{B}^{\prime } &=&\left[ \frac{\partial F_{B}\left( q_{0},\left| {\bf q}%
\right| \right) }{\partial q_{0}}+\frac{1}{\sqrt{1-4m^{2}/s}}\frac{\partial
F_{B}\left( q_{0},\left| {\bf q}\right| \right) }{\partial \left| {\bf q}%
\right| }\right] _{q_{0}\rightarrow 0,\left| {\bf q}\right| \rightarrow 
\sqrt{s/4-m^{2}}}
\end{eqnarray*}
To get a feeling of the underlying physics, note that for a constant form
factor $F_{B}=F_{B}^{\text{Const}}$, the absorptive parts are 
\begin{eqnarray*}
\func{Im}{\cal I}_{1}\left( s,x\right)  &=&\dfrac{\eta }{s}\dfrac{F_{B}^{%
\text{Const}}}{16\pi x}\ln \left[ \frac{1+\sqrt{1-\frac{4m^{2}}{s}}}{1-\sqrt{%
1-\frac{4m^{2}}{s}}}\right] \theta \left( s-4m^{2}\right)  \\
\func{Im}{\cal I}_{2}\left( s,x\right)  &=&-\dfrac{\eta }{s}\dfrac{F_{B}^{%
\text{Const}}}{16\pi x}\ln \left[ \frac{1+\sqrt{1-\frac{4m^{2}}{s\left(
1-x\right) }}}{1-\sqrt{1-\frac{4m^{2}}{s\left( 1-x\right) }}}\right] \theta
\left( s-\frac{4m^{2}}{1-x}\right) 
\end{eqnarray*}
The relative $-1$ factor is responsible for the destructive interference
when $x\rightarrow 0$. Explicitly, the SPL is 
\[
\func{Im}{\cal I}_{1}\left( s,x\right) +\func{Im}{\cal I}_{2}\left(
s,x\right) \stackrel{x\rightarrow 0}{\rightarrow }\frac{\eta }{s}\frac{%
F_{B}^{\text{Const}}}{16\pi }\frac{1}{\sqrt{1-4m^{2}/s}}
\]

{\bf iii- The dispersion integrals} are 
\begin{eqnarray}
{\cal I}_{1}\left( M^{2},x\right) &=&\frac{1}{\pi }\int_{4m^{2}}^{\infty }%
\frac{ds}{s-M^{2}}\func{Im}{\cal I}_{1}\left( s,x\right)  \label{PDMvert} \\
{\cal I}_{2}\left( M^{2},x\right) &=&\frac{1}{\pi }\int_{\frac{4m^{2}}{1-x}%
}^{\infty }\frac{ds}{s-M^{2}}\func{Im}{\cal I}_{2}\left( s,x\right)
\label{PDMobl}
\end{eqnarray}
We expect that the dispersive part will inherit the SPL properties of the
absorptive parts.

The oblique cuts are seen to be decreasing functions of $x$, and disappear
when $t^{2}\rightarrow 0$ ($x\rightarrow 1$). Further, in that limit ${\cal I%
}_{1}\left( s,x\right) \rightarrow {\cal I}\left( s\right) $, with ${\cal I}%
\left( s\right) $ the two-photon effective form factor (\ref{TwoPhotInt}).
At the decay rate level, the oblique cuts are really negligible because the
phase-space spectrum $\rho \left( x,a\right) $ is highly peaked around $x=1-a
$. In fact, independently of the form of ${\cal I}_{1}\left( s,x\right) $,
provided it is not too badly behaved, we can set ${\cal I}\left(
M^{2},x\right) \rightarrow {\cal I}_{1}\left( M^{2},1\right) $ in (\ref
{DiffRatepdm}) to get an estimation for the ratio as 
\begin{equation}
\frac{d\Gamma \left( p\text{-}Dm\rightarrow e^{+}e^{-}\gamma \right) /dx}{%
\Gamma \left( p\text{-}Dm\rightarrow \gamma \gamma \right) }=\frac{\alpha }{%
3\pi }\frac{\left| {\cal I}\left( M^{2},x\right) \right| ^{2}}{\left| {\cal I%
}\left( M^{2},1\right) \right| ^{2}}\rho \left( x,a\right) \approx \frac{%
\alpha }{3\pi }\rho \left( x,a\right)   \label{EstimDiff}
\end{equation}
with $a=4m_{e}^{2}/M^{2}$. This is a rather universal result: the highly
peaked spectrum wipes out the details of the dynamics, and the same ratio
holds for other pseudoscalar ($\pi ^{0}\rightarrow e^{+}e^{-}\gamma $, $%
K_{S}\rightarrow e^{+}e^{-}\gamma $,...). Integrated over $x$, this gives 
\begin{equation}
R_{Dm}^{th}\equiv \frac{\Gamma \left( p\text{-}Dm\rightarrow
e^{+}e^{-}\gamma \right) }{\Gamma \left( p\text{-}Dm\rightarrow \gamma
\gamma \right) }\sim \frac{\alpha }{3\pi }\left( 17.1\right) 
\label{EstimPDM}
\end{equation}
while for example $R_{\pi ^{0}}^{th}\sim \frac{\alpha }{3\pi }\left(
15.3\right) \sim 0.012$, in good agreement with experiments \cite{PDG}.

If we took some time to recall known stuff, it is to emphasize that the
present discussion of $p$-$Dm\rightarrow e^{+}e^{-}\gamma $ is really to be
taken as an illustration of some suppression effects. For the more
interesting orthopositronium decay, similar kinematical effects are
expected. However, comparing the respective phase-space photon spectrums
(figs. 6, 10), the suppression of oblique cuts is seen to be less effective
in orthopositronium.

We will now evaluate the contribution of each type of cuts for the case of
the Schr\"{o}dinger momentum wavefunction form factor (\ref{Schro1}). In
doing so, we will gain a better understanding of the analyticity restoration
at the level of the differential decay rate for $p$-$Dm\rightarrow
e^{+}e^{-}\gamma $.

\subsection{The vertical cuts reproduce standard approach results}

To precisely show what happens when the oblique cuts are forgotten, we
compute here the rate with the vertical cuts only. This is the result one
would obtain starting with (\ref{TH3}).

From the first cut imaginary part (\ref{Cut1Imaginary}), the decay amplitude
is found through an unsubstracted dispersion relation. The calculation is
very similar to that of the $p$-$Ps\rightarrow \gamma \gamma $ decay
amplitude, and we get 
\begin{equation}
{\cal I}_{1}\left( M^{2},x\right) =\frac{C\phi _{o}}{M^{3}}\frac{1}{x}\left[ 
\frac{2}{\pi }\arctan \frac{M}{2\gamma }\right]  \label{Cut1Real}
\end{equation}
Inserting the result for ${\cal I}_{1}$ into the expression of the
differential rate with $C^{2}=M/m^{2}$ and $\left| \phi _{o}\right|
^{2}=\alpha ^{3}m^{3}/8\pi $, we get: 
\[
\frac{d\Gamma \left( p\text{-}Dm\rightarrow \gamma e^{+}e^{-}\right) }{dx}=%
\frac{\alpha ^{6}m}{6\pi }\,\left( \frac{4m^{2}}{M^{2}}\right) \left| \frac{2%
}{\pi }\arctan \frac{M}{2\gamma }\right| ^{2}\frac{\rho \left( x,a\right) }{%
x^{2}} 
\]
Note that, independently of the value for $\gamma $, the photon spectrum is
always linear in $x$ for $x\approx 0$, due to the $1/x$ in (\ref{Cut1Real}).
This is an incorrect soft photon behaviour since Low's theorem requires a
cubic spectrum instead (the amplitude behaves as $x$, and an additional $x$
comes from phase-space). The $\gamma $-dependent corrections cancel in the
ratio 
\begin{equation}
\frac{d\Gamma \left( p\text{-}Dm\rightarrow \gamma e^{+}e^{-}\right) /dx}{%
\Gamma \left( p\text{-}Dm\rightarrow \gamma \gamma \right) }=\frac{\alpha }{%
3\pi }\frac{\rho \left( x,a\right) }{x^{2}}  \label{DiffVert}
\end{equation}
(compare with (\ref{EstimDiff})). For completeness, the total rate is
simply: 
\[
\frac{\Gamma \left( p\text{-}Dm\rightarrow \gamma e^{+}e^{-}\right) }{\Gamma
\left( p\text{-}Dm\rightarrow \gamma \gamma \right) }=\frac{\alpha }{3\pi }%
\left[ \frac{4}{3}\sqrt{1-a}\left( a-4\right) +2\ln \left( \frac{1+\sqrt{1-a}%
}{1-\sqrt{1-a}}\right) \right] 
\]
Numerically, 
\[
\frac{\Gamma \left( p\text{-}Dm\rightarrow \gamma e^{+}e^{-}\right) }{\Gamma
\left( p\text{-}Dm\rightarrow \gamma \gamma \right) }\approx \frac{\alpha }{%
3\pi }\left( 18.7\right) 
\]
to be compared with (\ref{EstimPDM}).

\subsection{The oblique cut contribution to the rate}

We have just seen that the vertical cuts suffice to reproduce the lowest
order decay rate. It is interesting to investigate how the oblique cuts can
restore the analytical behaviour of the spectrum without affecting this
lowest order evaluation.

Unfortunately, as mentioned earlier, the form factor to be used to compute
oblique cuts imaginary parts (and thereby their contributions to the decay
rate) is not the simple Schr\"{o}dinger momentum wavefunction, because the
energy-dependent part is not set to zero. To keep the discussion as concise
as possible, we will nevertheless use this Schr\"{o}dinger wavefunction,
i.e. 
\[
F_{B}\left( q_{0},{\bf q}\right) \equiv C\phi _{o}\frac{8\pi \gamma }{{\bf q}%
^{2}+\gamma ^{2}} 
\]
We could say that the standard approach neglect of subleading kinetic-energy
dependences is taken. In fact, since our purpose is only illustrative, the
present approximation is not very important. The cancellation leading to the
correct SPL can still be observed. Of course, the exact value of the oblique
cut contributions to the decay rate will not be exact, but this is not a
urgent question for now (the decay to $e^{+}e^{-}\gamma $ is about $1,5\%$
of the rate to $\gamma \gamma $ and the oblique cut corrections are at most
of a few percent, while $p$- $Dm$ has not even been observed yet!).

To be clear, basic principles (gauge invariance and analyticity) require the
introduction of oblique cuts, but practically, for low order evaluations,
their exact form is not relevant. On the contrary, a precise definition of
the form factor is necessary to tackle seriously the calculation of $o$-$%
Ps\rightarrow \gamma \gamma \gamma $, but this is left for future work
(because it may be more useful to uncover the remnants of oblique cuts in
standard NRQED expansions first).

When using the Schr\"{o}dinger form factor, the imaginary part of the
effective form factor has the constant SPL 
\begin{equation}
\func{Im}{\cal I}_{1}\left( s,x\right) +\func{Im}{\cal I}_{2}\left(
s,x\right) \stackrel{x\rightarrow 0}{\rightarrow }\frac{\eta }{s}\frac{C\phi
_{o}}{2}\frac{\gamma \left( s/4-\gamma ^{2}-m^{2}\right) }{\sqrt{1-\tfrac{%
4m^{2}}{s}}\left( \gamma ^{2}-m^{2}+\dfrac{s}{4}\right) ^{2}}  \nonumber
\end{equation}
We will now integrate the dispersion relation (\ref{PDMobl}). Obviously,
this integral is quite complicated and not very interesting for the present
purpose. Instead, we revert to numerical evaluation of ${\cal I}_{2}\left(
M^{2},x\right) $ as a function of $x$, and compare it to (\ref{Cut1Real}).

In the figure 12, we plot the vertical cuts contribution (dashed line),
normalized to get a constant as 
\[
\overline{{\cal I}_{1}}\left( M^{2},x\right) \equiv \left( \frac{xM^{3}}{%
C\phi _{o}}\right) {\cal I}_{1}\left( M^{2},x\right) =\frac{2}{\pi }\arctan 
\frac{M}{2\gamma } 
\]
The complete effective form factor, with the same normalization (solid line)
is 
\begin{eqnarray*}
\overline{{\cal I}}\left( M^{2},x\right) &\equiv &\left( \frac{xM^{3}}{C\phi
_{o}}\right) \left[ {\cal I}_{1}\left( M^{2},x\right) +{\cal I}_{2}\left(
M^{2},x\right) \right] \\
&=&\frac{2}{\pi }\arctan \frac{M}{2\gamma }+\frac{1}{\pi }\frac{xM^{3}}{%
C\phi _{o}}\int_{\frac{4m^{2}}{1-x}}^{+\infty }\frac{ds}{s-M^{2}}\func{Im}%
{\cal I}_{2}\left( s,x\right)
\end{eqnarray*}
The plots are drawn for $m=1,M=1.9975$ (i.e. $\gamma \approx 0.05$) and for $%
m=1,M=1.99975$ ($\gamma \approx 0.016$). Note that the physical value $%
\gamma \approx m\alpha /2$ corresponds to $\gamma \approx 0.0036$. For $%
x\rightarrow 0$, one can verify that $\overline{{\cal I}}\left(
M^{2},x\right) \rightarrow 0$.

The figures show that near zero the two cuts interfere destructively in
order to maintain a correct analytical behaviour for the whole amplitude.
Away from $x=0$, the oblique cut contributions are strongly suppressed
relatively to the vertical one, and this suppression increases as $\gamma $
decreases. As can be seen on the graph, it is typically for $x\lesssim
\gamma /m$ that the oblique cut contributes.

Therefore, we can summarize by giving a simple representation of the
different contributions to the amplitude. From the figure 12, one can see
that the behaviors of the vertical cuts, the oblique cuts and their
combination are quite precisely modelled as 
\[
{\cal I}_{1}\sim \frac{1}{x},\;{\cal I}_{2}\sim -\frac{\gamma /M}{x\left(
x+\gamma /M\right) }\Rightarrow {\cal I}={\cal I}_{1}+{\cal I}_{2}\sim
\left( \frac{1}{x+\gamma /M}\right) 
\]
As a consequence, the spectrum behaves as 
\begin{equation}
\frac{d\Gamma \left( p\text{-}Dm\rightarrow \gamma e^{+}e^{-}\right) }{dx}%
\sim \left| {\cal I}\right| ^{2}\rho \left( x,a\right) \sim x^{3}\left( 
\frac{1}{x+\gamma /M}\right) ^{2}  \label{ResultSpectrum}
\end{equation}
i.e. a linear spectrum when $\gamma \rightarrow 0$, and a $x^{3}$ spectrum
when $x$ is small. The effect of $\gamma \neq 0$ is therefore to soften the
photon spectrum and slightly reduce the total width. This behaviour is
exactly what we postulated in a previous work \cite{PreviousWrok}.\newline

In conclusion, the oblique cuts have a small contribution to the decay rate
comparatively to the vertical cuts. However, their presence is essential to
guarantee the analytical properties of the amplitude expected from Low's
theorem.

\subsection{Overview of $K_{S}\rightarrow e^{+}e^{-}\gamma $}

The restoration of analyticity by a cancellation among the vertical and
oblique cuts is presented in the context of the kaon decay $%
K_{S}^{0}\rightarrow e^{+}e^{-}\gamma $ via a charged pion loop (see \cite
{NonCstF} to \cite{chch}, and also \cite{TextBook2}). In this analysis, the
form factor describing the $K\rightarrow \pi \pi $ vertex is taken as a
constant. This decay is interesting since there is a close similarity
between this hadronic decay process and the present QED bound state decay.
The main conclusion is that even if charged particles are present in
intermediate states, the amplitude has to vanish in the soft photon limit,
as expected from the fact that the decays $K_{S}^{0}\rightarrow \gamma
\gamma ^{*}$ or $p$-$Dm\rightarrow \gamma \gamma ^{*}$ involve only neutral
external bosons. Further, this $K_{S}^{0}$ decay provides a physically
sensible process where one can analyze the constant form factor assumption,
since for such a loosely bound system as the $p$-$Dm$ a constant form factor
cannot be realistic.

Note that the modern approach to this decay channel is Chiral Perturbation
Theory, see for example \cite{Chiral}, \cite{SDChiral}.

\subsubsection{The pion loop model}

Details can be found in \cite{PreviousWrok}, so let us simply review the
main steps. The decay amplitude is modeled as a charged pion loop, to which
photons are attached either by the one-photon or the two-photon seagull
coupling (see figure 13). Gauge invariance is thereby enforced, but is
manifest only after the loop integration has been done using dimensional
regularization. At this stage, the amplitude reaches a form similar to (\ref
{TensorFactor}) 
\begin{eqnarray*}
{\cal M}\left( K_{S}^{0}\rightarrow \gamma e^{+}e^{-}\right)  &=&\dfrac{%
-2e^{3}}{\left( 4\pi \right) ^{2}}{\cal M}\left( K_{S}^{0}\rightarrow \pi
^{+}\pi ^{-}\right) \dfrac{1}{m_{\pi }^{2}}{\cal F}\left( M^{2},x\right)
\times  \\
&&\qquad \,\varepsilon _{\mu }^{*}\left( k\right) \left\{ \overline{u}\left(
p\right) \gamma _{\nu }v\left( p^{\prime }\right) \right\} \frac{g^{\mu \nu
}\left( k\cdot t\right) -t^{\mu }k^{\nu }}{t^{2}}
\end{eqnarray*}
with $t=p+p^{\prime }=P-k$, $m_{\pi }$ the pion mass, $x$ the reduced photon
energy $2k^{0}/M$ and $M$ the kaon mass. ${\cal F}\left( s,x\right) $ is the
Feynman parameter integral 
\begin{equation}
{\cal F}\left( s,x\right) =\int_{0}^{1}dy\int_{0}^{1-y}dz\frac{4zy}{%
1-4\left( a-b\right) zy+4by\left( y-1\right) +i\varepsilon }
\label{FPintegral}
\end{equation}
with the definitions $a=s/4m_{\pi }^{2},b=s\left( 1-x\right) /4m_{\pi }^{2}$
(in the $K$ rest-frame). This Feynman parameter integral plays exactly the
same role as ${\cal I}\left( M^{2},x\right) $, i.e. that of an effective
form factor (the only difference being that the momentum loop integration
has been done, and the imaginary part is obtained using the prescription $%
i\varepsilon $, instead of by cutting propagators, this is anyway strictly
equivalent). The differential rate is 
\[
\frac{d\Gamma \left( K_{S}^{0}\rightarrow e^{+}e^{-}\gamma \right) /dx}{%
\Gamma \left( K_{S}^{0}\rightarrow \pi ^{+}\pi ^{-}\right) }=\frac{\alpha
^{3}}{3\pi ^{3}}\frac{1}{a_{\pi }^{2}\sqrt{1-a_{\pi }}}\left| {\cal F}\left(
M^{2},x\right) \right| ^{2}\rho \left( x,a_{e}\right) 
\]
with $a_{\pi }=4m_{\pi }^{2}/M^{2},a_{e}=4m_{e}^{2}/M^{2}$, $m_{e}$ the
electron mass. The phase-space is given in (\ref{PhaseSpace}) with the same
bounds $x\in \left[ 0,1-a_{e}\right] $.

Using the same pion loop model, the two-photon decay rate is similarly
obtained, and 
\[
\frac{\Gamma \left( K_{S}^{0}\rightarrow \gamma \gamma \right) }{\Gamma
\left( K_{S}^{0}\rightarrow \pi ^{+}\pi ^{-}\right) }=\frac{\alpha ^{2}}{\pi
^{2}}\frac{1}{a_{\pi }^{2}\sqrt{1-a_{\pi }}}\left| {\cal F}\left(
M^{2},1\right) \right| ^{2}
\]
As previously mentioned (\ref{EstimDiff}), the ratio of the two
electromagnetic modes is approximately given by 
\[
\frac{d\Gamma \left( K_{S}^{0}\rightarrow e^{+}e^{-}\gamma \right) /dx}{%
\Gamma \left( K_{S}^{0}\rightarrow \gamma \gamma \right) }=\frac{\alpha }{%
3\pi }\frac{\left| {\cal F}\left( M^{2},x\right) \right| ^{2}}{\left| {\cal F%
}\left( M^{2},1\right) \right| ^{2}}\rho \left( x,a_{e}\right) \approx \frac{%
\alpha }{3\pi }\rho \left( x,a_{e}\right) 
\]
Anyway, as we will now discuss, the precise $x$ behavior is important
regarding the enforcement of the correct SPL.

\subsubsection{Dispersion Relations}

The effective form factor ${\cal F}\left( M^{2},x\right) $ is calculated
using dispersion relations \cite{TextBook2}, \cite{Sehgal}. Its imaginary
part consists of the vertical cuts (see figure 13) 
\[
\func{Im}{\cal F}_{1}\left( s,x\right) =\frac{2\pi m^{2}}{sx^{2}}\left\{ 
\frac{2m^{2}}{s}\ln \left[ \tfrac{1+\sqrt{1-\tfrac{4m^{2}}{s}}}{1-\sqrt{1-%
\tfrac{4m^{2}}{s}}}\right] -\left( 1-x\right) \sqrt{1-\tfrac{4m^{2}}{s}}%
\right\} \theta \left( s-4m^{2}\right) 
\]
and the oblique cuts 
\[
\func{Im}{\cal F}_{2}\left( s,x\right) =-\frac{2\pi m^{2}}{sx^{2}}\left\{ 
\frac{2m^{2}}{s}\ln \left[ \tfrac{1+\sqrt{1-\tfrac{4m^{2}}{s\left(
1-x\right) }}}{1-\sqrt{1-\tfrac{4m^{2}}{s\left( 1-x\right) }}}\right]
-\left( 1-x\right) \sqrt{1-\tfrac{4m^{2}}{s\left( 1-x\right) }}\right\}
\theta \left( s-\frac{4m^{2}}{1-x}\right) 
\]
(compare with $\func{Im}{\cal I}\left( s,x\right) $). Using an unsubstracted
dispersion relation 
\begin{equation}
{\cal F}\left( M^{2},x\right) =\func{Re}{\cal F}\left( M^{2},x\right) =\frac{%
1}{\pi }\int \frac{ds}{s-M^{2}}\func{Im}{\cal F}\left( s,x\right)
\label{DispReFab}
\end{equation}
for $M^{2}<4m_{\pi }^{2}$. The final expression for ${\cal F}$ can be
obtained by analytical continuation for any value of $M^{2}$%
\begin{equation}
{\cal F}\left( M^{2},x\right) =-\frac{1}{2\left( a-b\right) }+\frac{1}{%
\left( a-b\right) ^{2}}\left( \frac{1}{2}\left( f\left( a\right) -f\left(
b\right) \right) +b\left( g\left( a\right) -g\left( b\right) \right) \right)
\label{ResultfgFab}
\end{equation}
with 
\begin{eqnarray*}
f\left( x\right) &=&\left\{ 
\begin{array}{ll}
\arcsin ^{2}\left( \sqrt{x}\right) & 0<x<1 \\ 
-\left( \ln \left( \sqrt{x}+\sqrt{x-1}\right) -\frac{1}{2}i\pi \right) ^{2}
& x>1
\end{array}
\right. \\
g\left( x\right) &=&\left\{ 
\begin{array}{ll}
\sqrt{\frac{1-x}{x}}\arcsin \left( \sqrt{x}\right) & 0<x<1 \\ 
\sqrt{\frac{x-1}{x}}\left( \ln \left( \sqrt{x}+\sqrt{x-1}\right) -\frac{1}{2}%
i\pi \right) & x>1
\end{array}
\right.
\end{eqnarray*}
This result agrees with that of ${\cal O}\left( p^{4}\right) $ Chiral
Perturbation Theory \cite{Chiral}, \cite{SDChiral}, or with a direct
evaluation of the Feynman parameter integral (\ref{FPintegral}).

\subsubsection{Soft-Photon limits}

Individual contributions of each cut to the imaginary part are divergent as $%
\sim 1/x^{2}$. However, their combination is finite in the soft photon limit 
$x\rightarrow 0$. Indeed, applying L'Hospital's rule twice, we get 
\begin{equation}
\func{Im}{\cal F}_{1}\left( s,x\right) +\func{Im}{\cal F}_{2}\left(
s,x\right) \stackrel{x\rightarrow 0}{=}\frac{2\pi m^{4}}{s^{2}}\frac{1}{%
\sqrt{1-4m^{2}/s}}  \label{ImSoftLim}
\end{equation}
This constant SPL is not altered by the dispersion integrals, and the
effective form factor behaves as a constant for very low $x$%
\begin{equation}
{\cal F}\left( M^{2},x\right) \stackrel{x\rightarrow 0}{\rightarrow }-\frac{%
a_{\pi }}{4}+\frac{a_{\pi }^{2}}{4\left( a_{\pi }-1\right) }g\left( \frac{1}{%
a_{\pi }}\right)  \label{SoftLimit}
\end{equation}
(note that for $a_{\pi }<1$, the imaginary part of (\ref{SoftLimit}) is the
same as in (\ref{ImSoftLim})). Since there is a factor $k$ in the tensor
structure, the whole amplitude vanishes in the SPL, as expected from Low's
theorem.

Let us turn to the soft photon behaviour of the differential width. Since
the amplitude behave as $x$ for low $x$, the resulting spectrum is in $x^{3}$
($x^{2}$ from the squared amplitude and a $x$ from phase space). In other
words, if we had forgotten one cut, the resulting spectrum behaviour would
have been divergent as $1/x$ near zero instead of vanishing like $x^{3}$.
Compared to the paradimuonium and orthopositronium cases, it is even more
crucial not to forget oblique cuts, because of the IR divergence that would
be spuriously generated.

\subsubsection{Non-constant Form factor}

For a non-constant amplitude ${\cal M}\left( K_{S}^{0}\rightarrow \pi
^{+}\pi ^{-}\right) \equiv F\left( \left( q+k\right) ^{2},\left( q-t\right)
^{2}\right) $, structure-dependent terms must be supplemented to enforce
gauge invariance \cite{NonCstF}. By an analysis strictly parallel to the
orthopositronium one, the origin of the problem can be traced down to the
bremsstrahlung amplitudes 
\[
{\cal M}_{IB}^{\mu }\left( K_{S}^{0}\rightarrow \pi ^{+}\left( p_{1}\right)
\pi ^{-}\left( p_{2}\right) \gamma \left( k\right) \right) =F\left( \left(
p_{1}+k\right) ^{2},p_{2}^{2}\right) \frac{p_{1}^{\mu }}{p_{1}\cdot k}%
-F\left( p_{1}^{2},\left( p_{2}+k\right) ^{2}\right) \frac{p_{2}^{\mu }}{%
p_{2}\cdot k} 
\]
The effect of the structure terms is to restore both gauge invariance and
analyticity by accounting for the variations of $F$.\newline

In conclusion, the pseudoscalar decay $K_{S}^{0}\rightarrow e^{+}e^{-}\gamma 
$ is seen as an interesting system, being the low energy mesonic analogue of
dimuonium decay to $e^{+}e^{-}\gamma $. The dispersive techniques introduced
to deal with electromagnetic bound states are the standard tools used to
described $K_{S}^{0}\rightarrow e^{+}e^{-}\gamma $. Also, the same general
electrodynamical principles must hold. Namely, gauge invariance and
analyticity in the SPL are shown to be implemented by the same mechanism,
i.e. interference between imaginary part contributions. For non-constant
form factors, structure terms come into play to account for the variation of
the form factor, to preserve the cancellation among contributing cuts in the
SPL.

\section{Conclusions}

In this paper, we have used dispersion techniques to analyze positronium
decay amplitudes. In this way, emphasis is put on basic principles, i.e.
gauge invariance and soft photon limits. This provides interesting new
insights into bound state decay dynamics. Indeed, basic principles
implementation requires the non-perturbative treatment of some binding
energy effects. What our work has shown is that even if NRQED scaling
arguments are a very powerful calculational tool, the inherent breaking of
covariance hinders the manifestation of basic principles. In other words,
binding energy effects are usually perturbatively expanded along with
radiative corrections in the NRQED approach. The present work can therefore
be understood as complementary to NRQED, and may emphasize its underlying
model-dependent assumptions.

Some consequences are worth repeating here. First, no approximations were
needed to reach standard three-dimensional convolution-type decay amplitudes
(except of course the neglect of oblique cuts and structure-dependent
amplitudes). This provides a more concise (and consistent) scheme to deal
with spin structures and energy-dependences of the Bethe-Salpeter
wavefunction. As a by-product, we showed that some formulas usually quoted
in the literature are approximations, missing some of those ${\cal O}\left(
\alpha ^{2}\right) $ corrections they are meant to evaluate.

Also, our method emphasizes the fact that the exact integration of lowest
order decay amplitudes contains ${\cal O}\left( \alpha \right) $
corrections. This is not an artifact of our model. Rather, this is a
theoretical issue; a great deal of work in positronium decay calculation
being devoted to avoid double-counting. In our view, the most appropriate
formulation is to consider the corrections arising at the lowest order as
binding energy effects, and therefore to express them in terms of $\gamma $
(see (\ref{Improved2})). Further, this reordering of the non-relativistic
perturbation series may be much more adequate to deal with QCD bound state
like quarkonia, for which little is known about the wavefunction.

Concerning orthopositronium, a consistent picture of the decay process is
built. Gauge invariance and analyticity are at last correctly implemented.
Further, by exploiting the fact that some destructive interferences must
occur among the various contributions, bounds can be extracted on their
magnitudes. Unfortunately, only lower bounds are given, but one could
reasonably expect that those bounds serve as estimates, and conclude that
the present ${\cal O}\left( \alpha ^{2}\right) $ NRQED calculation is
complete. However, one should keep in mind that no proof of this conjecture
exists at present.

Let us make a comment about this optimistic conclusion. We have shown that
the decay process at lowest order in $\alpha $ requires the introduction of
both subleading contributions (oblique cuts) and additional
structure-dependent contributions. This requirement is to be understood at
the level of basic principles. In particular, gauge invariance is valid only
if structure terms are taken into account, i.e. gauge invariance does link
photons inside the bound state to radiating photons (emitted in the decay).
We have seen that NRQED proceeds by a perturbative reconstruction of the
Bethe-Salpeter loop. If gauge invariance is taken as a constraint in this
reconstruction, some contributions may be missed, because the loop is
ultimately not gauge invariant. Of course, since structure terms are invoked
to restore gauge invariance, and since they should be of order ${\cal O}%
\left( \gamma ^{3}\right) \sim {\cal O}\left( \alpha ^{3}\right) $, such
questions are not important practically at the present level of precision.
Anyway, it would be interesting to perform a thorough analysis of
gauge-dependent contributions arising in NRQED, in order to get an
independent estimate of structure-dependent contributions.

The central achievement of this work is to single out and characterize the
structure-dependent contributions as responsible for a breakdown of standard
NRQED at the ${\cal O}\left( \alpha ^{3}\right) $ level, or even sooner in
pathological cases. By breakdown is meant that considering only a two-body $%
e^{+}e^{-}$ Bethe-Salpeter wavefunction is no longer sufficient
(structure-dependent terms could in principle be obtained from a two-body to
three-body Green function). One should understand positronium calculations
as a two step process: the wavefunction is obtained from BS analyses, while
the decay process is calculated using NRQED. So it is really the
Bethe-Salpeter basis that is at stake, not the non-relativistic effective
theory. In our view, theoretical advances now require going beyond the
two-body Bethe-Salpeter approach. No answer to the orthopositronium lifetime
puzzle could be given before the completion of such progresses.

\qquad \newline

{\Large Acknowledgments: }C. S. and S. T. acknowledge financial supports
from FNRS (Belgium).

\section{Appendix}

This appendix contains the demonstration of the assertion that

\begin{quote}
{\it Dispersion relations constructed on the vertical cut imaginary parts of
the loop model reproduce standard convolution-type amplitudes.}
\end{quote}

We will particularize the discussion to the parapositronium decay into $%
2\gamma $, for which there are vertical cut contributions only. Let us
emphasize that the whole discussion of this section is readily extended to
any para- or orthopositronium vertical cut contributions to the decay
amplitude.

We first compute the imaginary part of (\ref{Loop1}) for an arbitrary
initial squared mass $P^{2}$. Considering the two possible cuts (figure 2),
we obtain $\func{Im}{\cal T}$ by replacing the two propagators on each side
of $\Gamma ^{\mu \nu }$ by delta functions 
\begin{eqnarray*}
\func{Im}{\cal T}\left( P^{2}\right) &\equiv &\func{Im}{\cal M}^{\mu \nu
}\left( p\text{-}Ps\rightarrow 2\gamma \right) \varepsilon _{1\mu
}^{*}\varepsilon _{2\nu }^{*} \\
&=&\int \frac{d^{4}q}{2\left( 2\pi \right) ^{2}}F_{B}\delta \left( \left( q-%
\frac{P}{2}\right) ^{2}-m^{2}\right) \delta \left( \left( q+\frac{P}{2}%
\right) ^{2}-m^{2}\right) \\
&&\quad \times Tr\left\{ \gamma _{5}\left( 
\!\not\!%
q-\frac{%
\!\not\!%
P}{2}+m\right) \Gamma ^{\mu \nu }\left( 
\!\not\!%
q+\frac{%
\!\not\!%
P}{2}+m\right) \right\} \varepsilon _{1\mu }^{*}\varepsilon _{2\nu }^{*}
\end{eqnarray*}
After a straightforward integration over $q^{0}$ and $\left| {\bf q}\right| $%
, with $P=\left( \sqrt{P^{2}},{\bf 0}\right) $, we reach 
\begin{eqnarray*}
\func{Im}{\cal T}\left( P^{2}\right) &=&\frac{1}{16\pi }\sqrt{1-\frac{4m^{2}%
}{P^{2}}}\theta \left( P^{2}-4m^{2}\right) \int \frac{d\Omega _{{\bf q}}}{%
4\pi }F_{B} \\
&&\quad \times Tr\left\{ \gamma _{5}\left( 
\!\not\!%
q-\frac{%
\!\not\!%
P}{2}+m\right) \Gamma ^{\mu \nu }\left( 
\!\not\!%
q+\frac{%
\!\not\!%
P}{2}+m\right) \right\} \varepsilon _{1\mu }^{*}\varepsilon _{2\nu }^{*}
\end{eqnarray*}
In the course of the derivation, the delta functions forced $q^{0}=0$ and $%
\left| {\bf q}\right| =\sqrt{P^{2}/4-m^{2}}$. In other words, the electron
momenta are 
\begin{equation}
\frac{1}{2}P\pm q=\left( \sqrt{\frac{P^{2}}{4}},\pm {\bf q}\right) \text{
with }\left( \frac{1}{2}P\pm q\right) ^{2}=\frac{P^{2}}{4}-\left| {\bf q}%
\right| ^{2}=m^{2}  \label{KinCut}
\end{equation}
This kinematics is to be understood in the trace evaluation. The angular
dependence arises from the relative orientations of ${\bf q}$ and the photon
momentum ${\bf l}_{1}$. Note also that the relation (\ref{KinCut}) cannot be
satisfied for the physical value $P^{2}=M^{2}<4m^{2}$. This is obvious since
the loop cannot have an imaginary part for the physical bound states, its
constituents being always off-shell. From the kinematics (\ref{KinCut}) one
can prove that the factors on both sides of $\Gamma ^{\mu \nu }$ are true
projectors, which serve to enforce gauge invariance in the expression 
\[
\left( 
\!\not\!%
q-\frac{%
\!\not\!%
P}{2}+m\right) \Gamma ^{\mu \nu }\left( 
\!\not\!%
q+\frac{1}{2}%
\!\not\!%
P+m\right) 
\]
Indeed, those two projectors play exactly the same role as external spinors
when demonstrating Ward identities.

The real part will now be calculated using an unsubstracted dispersion
relation 
\begin{equation}
\func{Re}{\cal T}\left( M^{2}\right) =\frac{P}{\pi }\int_{4m^{2}}^{+\infty }%
\frac{ds}{s-M^{2}}\func{Im}{\cal T}\left( s=P^{2}\right)  \label{ReTfi}
\end{equation}
where it is understood that $P^{2}$ should be replaced by $s$ everywhere,
i.e. scalar products that will appear when evaluating the trace should be
expressed with the kinematics defined for an initial energy $s$. Since $%
M^{2}<4m^{2}$, the principal part can be omitted and ${\cal T}\left(
M^{2}\right) =\func{Re}{\cal T}\left( M^{2}\right) $. Now let us write the
form factor in the general form 
\begin{equation}
F_{B}\equiv C\phi _{o}{\cal F}\left( {\bf q}^{2}\right) \left( {\bf q}%
^{2}+\gamma ^{2}\right) =C\phi _{o}{\cal F}\left( s/4-m^{2}\right) \cdot
\left( s-M^{2}\right) /4  \label{PostulatFb}
\end{equation}
with $\gamma ^{2}\equiv m^{2}-M^{2}/4$ and $\phi _{o}$ the bound state
wavefunction at zero separation. Then (\ref{ReTfi}) can be written as 
\begin{eqnarray*}
{\cal T}\left( M^{2}\right) &=&C\phi _{o}\int_{4m^{2}}^{+\infty }ds\int 
\frac{d\Omega _{{\bf q}}}{4\pi }{\cal F}\left( s/4-m^{2}\right) \frac{\sqrt{%
1-4m^{2}/s}}{64\pi ^{2}} \\
&&\quad \times Tr\left\{ \gamma _{5}\left( 
\!\not\!%
q-\frac{%
\!\not\!%
P}{2}+m\right) \Gamma ^{\mu \nu }\left( 
\!\not\!%
q+\frac{%
\!\not\!%
P}{2}+m\right) \right\} \varepsilon _{1\mu }^{*}\varepsilon _{2\nu }^{*}
\end{eqnarray*}

Let us transform the $s$ integral back into a $\left| {\bf q}\right| $
integral, keeping in mind the constraints obtained when extracting the
imaginary part. Using ${\bf q}^{2}=s/4-m^{2}$, $ds=8\left| {\bf q}\right|
d\left| {\bf q}\right| $, the decay amplitude dispersion integral is 
\[
{\cal T}\left( M^{2}\right) =\frac{C}{2}\phi _{o}\int \frac{d^{3}{\bf q}}{%
\left( 2\pi \right) ^{3}}\frac{{\cal F}\left( {\bf q}^{2}\right) }{\sqrt{%
P^{2}\left( {\bf q}\right) }}Tr\left\{ \gamma _{5}\left( 
\!\not\!%
q-\frac{%
\!\not\!%
P\left( {\bf q}\right) }{2}+m\right) \Gamma ^{\mu \nu }\left( 
\!\not\!%
q+\frac{%
\!\not\!%
P\left( {\bf q}\right) }{2}+m\right) \right\} \varepsilon _{1\mu
}^{*}\varepsilon _{2\nu }^{*} 
\]
where, as the notation suggests, it is understood that any $P^{2}$ appearing
in the amplitude must be replaced by $4\left| {\bf q}\right| ^{2}+4m^{2}$.
In particular, $\sqrt{P^{2}\left( {\bf q}\right) }$ can be replaced by $2E_{%
{\bf q}}$ with $E_{{\bf q}}=\sqrt{\left| {\bf q}\right| ^{2}+m^{2}}$. This
amounts to consider the scattering amplitude with incoming on-shell
electron-positron having momenta $\left( \frac{1}{2}P\left( {\bf q}\right)
\pm q\right) ^{2}=m^{2}$ (since $q^{0}=0$). Note the fact that $E_{{\bf q}%
}>M/2$, apparently the energy is not conserved. This is not surprising since
the present formula is a dispersion integral, done along the cut where $%
P^{2}\left( {\bf q}\right) >4m^{2}$. Finally, in view of the kinematics, we
introduce $k=\frac{1}{2}P\left( {\bf q}\right) +q$ and $k^{\prime }=\frac{1}{%
2}P\left( {\bf q}\right) -q$ (hence $E_{k}=E_{k^{\prime }}=E_{{\bf q}}$ and $%
{\bf k}=-{\bf k}^{\prime }={\bf q}$) to write the amplitude simply as 
\begin{equation}
{\cal T}\left( M^{2}\right) =\frac{C}{2}\int \frac{d^{3}{\bf k}}{\left( 2\pi
\right) ^{3}2E_{{\bf k}}}\left[ \phi _{o}{\cal F}\left( {\bf k}^{2}\right)
\right] Tr\left\{ \gamma _{5}\left( -%
\!\not\!%
k^{\prime }+m\right) \Gamma ^{\mu \nu }\left( k,k^{\prime },l_{1}\right)
\left( 
\!\not\!%
k+m\right) \right\} \varepsilon _{1\mu }^{*}\varepsilon _{2\nu }^{*}
\label{StdDec}
\end{equation}
where $\Gamma ^{\mu \nu }\left( k,k^{\prime },l_{1}\right) $ is the
amplitude for on-shell $e^{-}\left( k\right) e^{+}\left( k^{\prime }\right) $
scattering into $2\gamma $. Gauge invariance is present due to the two
projectors, well defined since $k^{2}=k^{\prime 2}=m^{2}$. This ends our
demonstration, and ${\cal T}\left( M^{2}\right) ={\cal M}\left( p\text{-}%
Ps\rightarrow \gamma \gamma \right) $.


\begin{thebibliography}{99}
\bibitem{Discovery}  M. Deutsch, Phys. Rev. {\bf 82}, 455 (1951).

\bibitem{Experm}  A. Al-Ramadhan, D. Gidley, Phys. Rev. Lett. {\bf 72}, 1632
(1994).

\bibitem{Tokyo}  S. Asai, S. Orito, N. Shinohara, Phys. Lett. {\bf B357},
475 (1995); O. Jinnouchi, S. Asai, T. Kobayashi, {\it hep-ex/0011011}.

\bibitem{AnnArborGas}  C. Westbrook, D. Gidley, R. Conti, A. Rich, Phys.
Rev. {\bf A40}, 5489 (1989).

\bibitem{AnnArborVac}  J. Nico, D. Gidley, A. Rich, P. Zitzewitz, Phys. Rev.
Lett. {\bf 65}, 1344 (1990).

\bibitem{History}  J. A. Wheeler, Ann. N. Y. Acad. Sci. {\bf 48}, 219
(1946); J. Pirenne, Arch. Sci. Phys. Nat. {\bf 29}, 265 (1947).

\bibitem{OrePowell}  A. Ore and J. L. Powell, Phys. Rev. {\bf 75}, 1696
(1949).

\bibitem{Harris}  I. Harris and L. Brown, Phys. Rev. {\bf 105}, 1656 (1957).

\bibitem{Lepage79}  W. E. Caswell, G. P. Lepage, Phys. Rev. {\bf A20}, 36
(1979)

\bibitem{Tomozawa}  Y. Tomozawa, Ann. Phys. {\bf 128}, 463 (1980)

\bibitem{Adkins}  G. Adkins, Ann. Phys. {\bf 146}, 78 (1983).

\bibitem{RecentCorr}  A. Czarneski, K. Melnikov, A. Yelkhovsky, Phys. Rev.
Lett. {\bf 83}, 1135 (1999); eprint {\it hep-ph/9910488}; G. Adkins, R.
Fell, J. Sapirstein, Phys. Rev. Lett. {\bf 84}, 5086 (2000); B. Kniehl, A.
Penin, Phys. Rev. Lett. {\bf 85}, 1210 (2000); K. Melnikov, A. Yelkhovsky,
Phys. Rev. {\bf D62}, 116003 (2000).

\bibitem{LargeRelCorr}  W. E. Caswell, G. P. Lepage, J. Sapirstein Phys.
Rev. Lett. {\bf 38}, 488 (1977).

\bibitem{LargeRelCorr2}  I.B. Khriplovich, A.I. Milstein, J. Exp. Theor.
Phys. {\bf 79}, 379 (1994).

\bibitem{BS2}  V. Antonelli, V. Ivanchenko, E. Kuraev, V. Laliena, Eur.
Phys. J. {\bf C5}, 535 (1998), V. Antonelli,{\it \ Int. Work. on Hadronic
Atoms and Positronium in the S.M.}, Dubna, 26-31 May 1998.

\bibitem{MultiPhot}  G. Adkins, F. Brown, Phys. Rev. {\bf A28}, 1164 (1983);
G. P. Lepage, P. B. Mackenzie, K. H. Streng, P. M. Zerwas, Phys. Rev. {\bf %
A28}, 3090 (1983)

\bibitem{NRQED}  W.E.\ Caswell and G.P. Lepage, Phys. Lett. {\bf B167}, 437
(1986); P. Labelle, MRST Meeting 1992 (CLNS-92-1161).

\bibitem{Low}  F. E. Low, Phys. Rev. {\bf 110}, 974 (1958).

\bibitem{PreviousWrok}  G. Lopez Castro, J. Pestieau and C. Smith, {\it %
hep-ph/0004209}, G. Lopez Castro, J. Pestieau, C. Smith and S. Trine, {\it %
hep-ph/0006016 }and{\it \ hep-ph/0006018}.

\bibitem{TextBook1}  M. Peskin, D. Schroeder, {\it An Introduction to
Quantum Field Theory}, Addison-Wesley (1995); O. Nachtmann, {\it Elementary
Particle Physics : Concepts and Phenomena, }Springer-Verlag (1990)

\bibitem{TextBook2}  K. Nishijima, {\it Fields and Particles : Field Theory
and Dispersion Relations}, Benjamin N.Y. (1969).

\bibitem{Kniehl}  B. Kniehl, Acta Phys. Polon. {\bf B27}, 3631 (1996) ({\it %
hep-ph/9607255}).

\bibitem{TextBook3}  F. Gross, ''{\it Relativistic quantum Mechanics and
Field Theory}'', Wiley (1993).

\bibitem{BarbRem}  R. Barbieri and E. Remiddi, Nucl. Phys. {\bf B141}, 413
(1978).

\bibitem{Wallace}  D.R. Phillips, S.J. Wallace, Phys. Rev. {\bf C54}, 507
(1996)

\bibitem{RadCorr}  L. M. Brown, R. P. Feynman, Phys. Rev. {\bf 85}, 231
(1952)

\bibitem{Dimuonium}  U. Jentschura, G. Soff, V. Ivanov and S. Karshenboim,
Phys. Rev. {\bf A56}, 4483 (1997); U. Jentschura, G. Soff, V. Ivanov, S.
Karscenboim, {\it hep-ph/9706401}.

\bibitem{NonCstF}  H. Chew, Phys. Rev. {\bf 123}, 377 (1961); J. Pestieau,
Phys. Rev. {\bf 160}, 1555 (1967).

\bibitem{Chiral}  G. Ecker, A. Pich, E. de Rafael, Nucl. Phys. {\bf B303},
665 (1988)

\bibitem{SDChiral}  J. Bijnens, G. Ecker and J. Gasser, Nucl. Phys. {\bf B396%
}, 81 (1993); G. D'Ambrosio, G. Ecker, G. Isidori, H. Neufeld, 2$^{nd}$
DAPHNE Physics Handbook:253-313 ({\it hep-ph/9411439}).

\bibitem{Sehgal}  L.M. Sehgal, Phys. Rev. {\bf D7}, 3303 (1973).

\bibitem{Bergstr}  L. Bergstr\"{o}m, G. Hulth, Nucl. Phys. {\bf B259}, 137
(1985); J. L. Lucio M., J. Pestieau, Phys. Rev. {\bf D42}, 3253 (1990).

\bibitem{chch}  L. L. Chau, H. Y. Cheng, Phys. Lett. {\bf B195}, 275 (1987).

\bibitem{PDG}  C. Caso {\it et al}, Review of Particle Physics, Eur. Phys.
J. {\bf C3}, 1, 1998.
\end{thebibliography}
\end{document}